\documentclass[10pt,journal,compsoc]{IEEEtran}

\ifCLASSOPTIONcompsoc
  \usepackage[nocompress]{cite}
\else
  \usepackage{cite}
\fi

\usepackage{balance}
\usepackage{amsmath,amsthm,amssymb,amsfonts,stmaryrd,mathabx,manfnt,latexsym}
\usepackage{paralist}
\usepackage{url}
\usepackage{listings}
\usepackage{color}
\usepackage{multirow}

\usepackage{graphicx}
\usepackage{stfloats}

\ifCLASSOPTIONcompsoc
  \usepackage[caption=false,font=footnotesize,labelfont=sf,textfont=sf]{subfig}
\else
  \usepackage[caption=false,font=footnotesize]{subfig}
\fi

\usepackage{array}
\usepackage{algorithm} 
\usepackage{tikz}
\usetikzlibrary{positioning}
\usepackage{algpseudocode}

\theoremstyle{plain}
\newtheorem{lemma}{Lemma}
\newtheorem{theorem}[lemma]{Theorem}
\newtheorem{proposition}[lemma]{Proposition}

\theoremstyle{definition}

\newtheorem{definition}{Definition}[section]

\begin{document}

\title{Energy Conscious Dynamic Window Scheduling\\ of Chip Multiprocessors}
\author{Matthew Michel and Hyunyoung Lee~\IEEEmembership{Member,~IEEE}}

\IEEEtitleabstractindextext{
\begin{abstract}
The need to develop systems that exploit multi and many-core architectures to
reduce wasteful heat generation is of utmost importance in compute-intensive applications.
We propose an energy-conscious approach to multicore scheduling known as non-preemptive dynamic window (NPDW) scheduling that achieves effective load and temperature balancing
over chip multiprocessors.

NPDW utilizes the concept of dynamic time windows to accumulate tasks and find an
optimal stable matching between accumulated tasks and available processor cores using
a modified Gale-Shapely algorithm. The metrics of window and matching performance are 
defined to create a dynamic window heuristic to determine the next time window size 
based on the current and previous window sizes.

Based on derived formulation and experimental results, we show 
that our NPDW scheduler is able to distribute the computational and thermal load   
throughout the processors in a multicore environment better than baseline schedulers. 
We believe that within multicore compute applications requiring temperature and 
energy-conscious system design, our scheduler may be employed to effectively 
disperse system load and prevent excess core heating.
\end{abstract}

\begin{IEEEkeywords}
Chip multiprocessors, Energy-aware systems, Temperature-aware design, 
Dynamically-scheduled, Load balancing and task assignment
\end{IEEEkeywords}}

\maketitle

\IEEEraisesectionheading{\section{Introduction}\label{introduction}}

\IEEEPARstart{T}{he} advent of chip multiprocessors (CMPs) has enabled great improvements in processor performance. Historically, processor performance has been improved by increasing processor clock speed along with the number of transistors on a computer chip~\cite{9128075}. However, due to the power, heat, and size limitations of physical hardware, the rate of processor improvements from these traditional methods has significantly decreased in the past several decades~\cite{9128075}. Chip multiprocessors present a design architecture in which several processing units are placed on a single electrical chip enabling concurrent, localized execution~\cite{1385946}.
The concurrent execution and scheduling of threads in this system is known as chip multithreading. As a consequence of the development of chip multithreading, improvement and redevelopment of task scheduling techniques is a necessary ongoing research topic. 
In particular, effective utilization and scheduling of CPU cores such that load 
and heat generation are 
equally distributed is an area of great interest. Recent research has indicated that modern day schedulers of widely used systems such as the Linux Kernel continue to suffer performance issues due to multicore hardware architectures~\cite{decade-wasted-cores}.

We propose a scheduling algorithm that 
equitably distributes task load among CPU cores to maximize system throughput and thermal distribution. Cores will be treated as interdependent entities who mutually work together to achieve optimal cumulative performance and thermal load balancing. In order to distribute scheduled tasks (or processes) among CPU cores, we propose  
an efficient matching scheme based on the Gale-Shapely algorithm~\cite{1989827} that 
assigns non-preemptive, non-conflicting processes to processor cores within a dynamic time frame. 
The matching scheme in our system is represented
as a bipartite graph consisting of processes (labeled as consumers) and processor cores. 
The active consumers are continuously changing as processes arrive and are scheduled to be executed.
The set of currently available processor cores is continuously changing as the processor cores complete the assigned tasks. Each consumer has an associated weighted profile which gives quantitative details about its processing demand. This weighted profile 
is utilized to match the consumer to a processor.
Additionally, we include a credit system 
in order to achieve a fair scheduling in the view of processes and ensure approximately 
equal CPU distribution among the cores. 
The overall processor throughput along with matching performance via average waiting time 
is used to adapt the scheduler in order to achieve greater performance. 

The main contribution of our research presented in this paper is to investigate a task scheduling 
scheme that concerns throughput, power consumption, and heat dissipation which are 
indispensable concerns for modern day computing systems. Furthermore, we apply the concept 
of bipartite matching and dynamic windowing to task scheduling 
in a chip multiprocessor and chip multithreaded hardware architecture. Within this application, 
we develop a matching algorithm based on the Gale-Shapely algorithm~\cite{1989827}.

The rest of the paper is structured as follows. In the next section we address  
related work in dynamic approaches to processor 
scheduling as well as matching. In Section~\ref{system}, the system model is proposed. 
An in-depth explanation of our algorithm with pseudocode is provided in Section~\ref{algorithm}. 
Section~\ref{experiments} discusses experimental data and the simulation results of our 
algorithm, and Section~\ref{conclusions} is the concluding body of our paper.

\section{Related Work} \label{related}

With regards to minimizing power consumption in resource scheduling, the authors 
in~\cite{8639212} detail a heuristic for maximizing power efficiency in both homogeneous 
and heterogeneous chip multiprocessors as measured through energy-delay product. 
The authors in~\cite{6904264} 
approach the question of scheduling with priority queues and duplication of tasks, 
seeking to shorten length of task scheduling as well as maximize processor utilization. 
The authors in~\cite{3358603} distinguish and provide scheduling implementations for both preemptive and non-preemptive scheduling. 

Thermal management and equitable distribution of heat throughout multicore systems
is an area of interest and needed research. The authors in~\cite{temp-sched-decisions}
approach this problem by considering a temperature based policy driven by a linear 
combination between core utilization and temperature.
This scheduler runs alongside the default system scheduler, only being activated 
when approaching the maximum allowable system temperature. Within the thermal 
policy, weights associated with the temperature parameter increase with temperature
leading the scheduler to favor lower temperature cores for process execution. 
In~\cite{proactive-thermal-aware}, authors propose the Simple Time Derivative 
scheduler, a scheduler that utilizes core temperature derivatives over time to
predict future temperature. In particular, once a certain threshold of the 
temperature derivative is met, 
throttling of ``hot'' processes occurs where hot processes are noted as containing heavy 
usage of both integer and floating point arithmetic.
Authors in
~\cite{dynamic-thermal-management} implemented the ThreshHot scheduler seeking to
reduce violations of thermal thresholds. The scheduler hinges on the idea that
when scheduling two jobs, scheduling the hottest job first will result in a lower 
overall temperature after the cooler second job is scheduled 
than if the hottest 
job was scheduled second. This idea is implemented in the scheduler by choosing
the hottest job that is believed to not violate the thermal threshold whenever
a new job is available for scheduling. Decreases in thermal violations were 
observed and the researchers saw increases in overall system throughput due to 
the lack of throttling.
Lastly, in~\cite{dynamic-time-slice}, 
authors utilize a dynamically changing time slice in order to regulate thermal
temperature alongside fairness. Within the scheduler, each process was assigned
its own time slice. If a process was observed to be excessively heating the 
CPU, the time slice of the process would be decreased in order to lessen 
the amount of heat it contributes to the system.  

In multithreaded applications, the execution of the program is not guaranteed to be completed 
with multiple concurrent threads in its entirety. Some portion of the program's execution will 
most likely consist of a single running thread. The authors in~\cite{5715064} acknowledge 
this distinction and separate task scheduling based upon whether or not threads are 
executing concurrently, ultimately increasing overall throughput.

Based on related research in the area of task scheduling in chip multiprocessors, we observed 
large amounts of focus on heavily parallelized tasks and scheduling arrangements of subtasks. 
While thread-level parallelism offers highly beneficial exploitation of the chip multiprocessor 
architecture and is deserving of large amounts of research effort, it must be acknowledged 
that many tasks scheduled for execution on personal computing environments are entirely 
unrelated. Consequently, scheduling algorithms written for the sole purpose of scheduling 
tasks with high parallelization and dependency may not yield the same benefit when a larger 
queue of unrelated tasks are considered. This is substantiated by the finding and algorithmic 
design used in~\cite{5715064} in which individually handling ``serial phases'' and 
``parallel phases'' increased processor throughput. Our approach to task scheduling places 
strong consideration on these independent tasks, considers the time-frame and relative 
predictability of when tasks may be scheduled, and provides flexible adaptation to different 
arrival patterns of tasks via a concept we define as dynamic windowing
in time based on core utilization and temperature.

\section{The System Model} \label{system}

We provide a holistic overview of the environment in which our algorithm executes as well as 
system architecture and requirements.
We also define input parameters and the credit system.

\subsection{Formal Problem Representation} \label{formal}
Our system model seeks to efficiently identify a bipartite matching within a forest of nodes, 
$G = (U, V)$ where $U$ and $V$ are disjoint sets such that $U \cap V = \emptyset$. 
A modified Gale-Shapely algorithm~\cite{1989827} implemented in Section~\ref{algorithm} 
is used to obtain the matching.
Each $u \in U$ is an independent task that needs to be scheduled for execution on the local 
system processor. The set $U$ dynamically changes as processes arrive.
Processes may reappear after being executed and previously unseen processes may also 
arrive. Each $v \in V$ 
represents a unique processor core in a chip multiprocessor, which is currently available for a 
task to be scheduled to run on. The set $V$ of processor cores dynamically changes as 
processes are scheduled to run on the processors 
and the processors complete the assigned tasks. 
Task arrival works under the assumption that once a task yields the CPU for I/O, the recurring 
task will be treated as a new instance.
The matching of $G$ is the set $E$ of edges between $u\in U$ and $v\in V$ such that
edge $(u,v) \in E$ represents a task $u$ being assigned to a specific processor core $v$ for 
execution. Each $u \in U$ and $v \in V$ has a maximum degree of one. Our solution to the 
proposed problem is twofold. We first seek to identify effective matchings of $(u, v) \in E$ 
based on characteristics of the nodes in $U$ and $V$ at a given point in time and return the 
matching set $E$ from our algorithm. Additionally, we seek to find a sufficient window of time 
to collect elements (consumer tasks) in $U$ and processor cores in $V$ for the matching. 
This period of time is known as the \textit{accumulation window}.

\subsection{Hardware Requirements and Assumptions} \label{hardware}
While the task scheduling algorithm is intended to be processor agnostic and applicable to 
popular commercial and personal central processing units, the needed parameters of our 
algorithm include necessary processor requirements that are only present in modern 
architecture designs. 
First, we define a core as the smallest unit of independent CPU execution in a processor. 
Alternate sources may refer to this as a CPU thread or virtual core. This definition of the term 
will be used throughout the remainder of the paper. The first requirement of our processor is 
that it implements a chip multiprocessor architecture which is a near universal standard in 
modern day computers. Our algorithm focuses on assignment of tasks to processor cores and 
would yield no benefit in any single-core platform. Additionally, it is important that the cores are 
homogeneous indicating that all processor cores are of the same model and specification. 
The matching algorithm assumes that cores are identical and offer the same capacity for 
performance. 
The last 
requirement is that the CPU provides individual core temperature data as in~\cite{7000001}.

The assembly analysis step of our algorithm assumes the instruction set of an x86-64 architecture.  
Within this x86-64 architecture, we make categorizations of opcode instruction types with 
different weightings to calculate instruction time and power cost.

\subsection{Parameters of Interest} \label{parameters}
Our system creates a weighted average of various input parameters for each provided task in 
order to assign the task to a processor core. Tasks are assigned in such a way to balance 
processor throughput with equalizing workload
and scheduling fairness. Analyzing the process prior to matching is useful in determining the 
process' expected runtime, power consumption, and overall burden on the processor. These 
parameters allow the algorithm to assign the process to a matching processor core at an 
opportune time.
First, there are four parameters required of the tasks.
\begin{compactenum}
\item[$p_1$:]
Predicted average instruction time cost categorization. 
\item[$p_2$:]
Predicted average instructional power consumption categorization. 
\item[$p_3$:] 
Optional execution deadline; 
may be used to allow other processes to be scheduled first if necessary;
is integrated into the task credit system defined in Section~\ref{credit}.
\item[$p_c$:]
Current credit of the task; 
calculated and maintained by the task credit system 
defined in Section~\ref{credit}.
\end{compactenum}
Metrics $p_1$ and $p_2$ are created internally within our system based on
assembly instructions opcode analysis. 
Our assembly analyzer (Section~\ref{static}) scans and quantifies the types and number of 
opcode instructions to be used, similarly to  
previous findings (e.g.,~\cite{LEITE20142260}) and the measured instruction table in~\cite{7000000}.
Then, weighted averages of the instruction types with their associated time costs (for $p_1$) 
and power consumption (for $p_2$) are calculated and each parameter is assigned a value 
$\in \{1,2,3\}$ corresponding to the categorization of \textit{low}, \textit{moderate}, or \textit{high}. 

Categorizations are used instead of quantitative estimates for several reasons. Firstly, numeric 
estimation of a program's instruction time and power consumption cost is rather unpredictable 
and highly processor contingent. Even slightly differing 
external factors may result in significant disparities in both execution time and power consumption for the same process. Additionally, the real-time nature of the assembly analysis
requires near instantaneous generation of these parameters. Predictions with too high precision 
may become an unnecessary bottleneck to the entire scheduling algorithm. Finally, the purpose 
of collecting these parameters is to compare process instruction time and power cost relative to 
other processes.
For instance, if we consider two potential schedules $a$ and $b$ for a free processor core such 
that $a$ is deemed to be more optimal than $b$, then $a$ will always be the chosen schedule. 
Therefore, high precision instruction time and power consumption estimates are not necessary 
for our use case.

We represent these four parameters together with the process (task) ID $p_0$ by a 
tuple with five components: $(p_0, p_1, p_2, p_3, p_c)$.
Then, $P$ denotes the set  
containing those five-component tuples of all tasks currently in the waiting queue.
That is, each element $p\in P$ is a five-component tuple for a process currently in the task 
queue whose ID is stored in $p_0$, and $p_1, p_2, p_3,$ and $p_c$ are its parameters.

Next, we address the parameters provided by the processor cores to the system. These 
parameters are characteristic data of the current state of the CPU core and are obtained by 
system calls to the operating system. 
\begin{compactenum}
\item[$c_1$:] Core speed (unit: MHz, tool to obtain: /proc/cpuinfo); 
on average the same per core (as the cores are homogeneous); 
core speeds can differ when it heats up
to the point where the hardware enacts thermal throttling and reduces the internal clock speed 
as a self-preservation mechanism~\cite{7000001} or when the core is underutilized and 
decreases speed to save power. 
\item[$c_2$:] 
Core load (unit: $\%$, tool to obtain: /proc/stat);
the percentage of the core being utilized for processing within a given time window. 
\item[$c_3$:]
Core temperature (unit: $^\circ$C, tool to obtain: Linux Hwmon); 
is obtained and categorized in one of three temperature states -- low, moderate, or high 
(the corresponding value stored as $\{1,2,3\}$). 
\item[$c_{msp}$:]
The maximum processor speed in the system (unit: MHz, tool to obtain: /proc/cpuinfo);
the same for all cores in the system.
\end{compactenum}
We represent these four parameters together with the processor core ID $c_0$ by a 
tuple with five components: $(c_0, c_1, c_2, c_3, c_{msp})$.
Then, $C$ denotes the set  
containing those five-component tuples of available cores in the system.

\subsection{Time Window Approach} \label{time}
An approach we refer to as dynamic windowing 
is utilized in order to determine a period of wait time that allows enough tasks to arrive
and enough cores to become available to form an effective task-processor matching. 
We assume some measurable distribution of arrival of tasks to the system and in the 
availability of CPU cores. Different application software has variability in the types of processes 
scheduled including the frequency of their execution, their power consumption, execution time 
complexity, and other process characteristics. Additionally, CPU cores may exhibit high or low 
utilization and/or temperature. Altering the window of scheduling time grants our system flexibility 
in accounting for these differences and to achieve a window that leads to a 
more effective scheduling. This window is referred to as the \textit{accumulation window}. 
At the end of the window a matching followed by execution on the processor occurs. 
The time devoted to carrying out the execution of a matching is referred to as 
the \textit{execution window}. Ultimately, once the optimal window of time for collecting 
processes and cores to be scheduled is determined for an instance of our algorithm, the 
fluctuations in window size will decrease while still supporting dynamic adaptation for 
future changes in scheduling behavior.

\subsection{Credit System} \label{credit}
A credit system is established in order to balance the execution of processes and to prevent 
any singular process from routinely hogging processor resources. Each 
incoming task is assigned a base number of credits. The tasks are rewarded credits for allowing 
their execution 
to be delayed during the matching process. Tasks that require immediate or near-immediate 
execution will have to expend an amount of credit to do so. 
When performing a matching between CPU cores and tasks, task credit is weighted in the 
matching process. This causes tasks with high credit to be more likely scheduled than tasks 
with lower credit if there are not enough cores to schedule all tasks.

In establishing a credit system there are two functions we must provide: A function for credit 
gain (reward system for waiting to be scheduled), and a function for credit spending in which 
tasks spend their own credit to be scheduled. Within this credit system, the range of valid credit 
is $[0, \infty)$. While the lower bound of credit is zero 
(tasks may not enter credit debt), 
the upper bound of credit gain is limitless, although in practice a task is not able to specify an 
extraordinarily long tentative deadline due to the finite length of the accumulation windows and 
the maximum limit on how many rounds of scheduling a task may be passed set by the system. 

\begin{definition}[Credit Accumulation Function] \label{credit-acc-fn}
We define credit accumulation recursively as
$$f_0 = b \enspace \text{ and } \enspace f_n(\Delta t) = c\Delta t + f_{n-1}$$
where $b$ is the base credit all tasks begin with, $c$ represents a constant of credit gained
per unit time, $n$ is the number of accumulation windows the task has waited in the queue since 
its arrival in the system, and $\Delta t$ represents the elapsed time the task has been in the 
set $U$ since the start of the $n$th accumulation time window.
The most recent calculated value of $f$ is assigned to $p_c$. 
\end{definition}

\begin{definition}[Credit Spending Function] \label{credit-spending}
The amount of credit a task is charged immediately prior to execution is
\begin{equation}
\label{eqn:credit_spending}
g (p_c, U) = p_c (1 - {1}/{|U|})
\end{equation}
where $p_c$ is the credit of the task to be scheduled.
\end{definition}
\begin{proposition}[Fairness of the Credit Spending Function] \label{credit-spending-fairness}
The credit spending function described in Definition~\ref{credit-spending} ensures equitable 
credit charging of all tasks based on proportional credit charges for execution.

\begin{proof}
Using the credit spending function~(\ref{eqn:credit_spending}), a task expends a variable 
percentage rate of credit. For instance, a singular task scheduled alone will be charged $0$ 
credit since it does not have to bypass any tasks to execute. However, as the scheduler 
becomes loaded with an increased number of tasks, an increasingly scaled percentage of credit 
will be charged to execute. Since credit itself factors into the decision of the scheduler to execute 
a task, tasks with higher credit will incur a greater credit cost. However, since the percentile rate 
for any scheduled task to execute at a period of time is the same, our scheduler still maintains 
this aspect of impartiality.
\end{proof} 
\end{proposition}

\section{The Algorithm} \label{algorithm}
Our algorithm consists of four parts: First, static analysis of the assembly code of the task needs 
to be performed to determine the two parameters $p_1$ and $p_2$ before the task is considered 
as a consumer process.
Second, task-to-core and core-to-task preference lists need to be generated
which will be followed by the third step of matching.  
The fourth part is the dynamic windowing based on the performance measures of throughput 
and power consumption as well as the level of consumer satisfaction.
      
\subsection{Static Assembly Analysis} \label{static}
Prior to a process being received as a consumer in our matching algorithm, a programmatic 
analysis of the process' assembly code must occur to generate the proper parameters to be 
used as input for the matching. These parameters categorize the process by average instruction 
time cost $p_1$ and average instruction power consumption $p_2$.

Each category of instructions is assigned a weighting relative to the performance of the other instructions.
The assignment is based on empirical measurements via benchmarking on the host machine regarding general purpose and SIMD Extension instructions made by Intel~\cite{7000002} for instruction execution. 

\vspace*{-3pt}
\begin{table}[h]
\centering
\caption{\label{tab:opcode-time-weight} Relative Opcode Time Weightings}
\begin{tabular}{|c|c|c|c|}
\hline
 \multicolumn{2}{|c|}{\textbf{Category}} & \textbf{Example Opcodes} & \textbf{Weight} \\
\hline
\multirow{2}{*}{C1} & Addition, Subtraction, and & addl, subl, & \multirow{2}{*}{1.00} \\
 & Integer Multiplication & mull, etc. &\\
\hline
C2 & Integer Division & idivl & 1.42 \\
\hline
C3 & Floating Point Multiplication & mulsd & 1.03 \\
\hline
C4 & Floating Point Division & divsd & 1.66 \\
\hline
C5 & Heap Memory Operations & $\_\_$malloc & 12.30 \\
\hline
C6 & Stack Memory Operations & movl, leal, etc. & $\sim$1\\
\hline
C7 & Control Instructions & jge, jmp, jle, etc. & $\sim$1\\
\hline
C0 & (no category) & ret & 0\\
\hline
\end{tabular}
\end{table}

With regards to power consumption,
we assume that required energy per instruction can be approximated. We 
utilize ratios established based upon measurements in~\cite{LEITE20142260} to assume the 
relative cost between the different power consumption categorizations. The ratios used are 
based on quantitative measurements taken on a CPU with a frequency of $2.4$ GHz. 

\begin{table}[h]
\centering
\caption{\label{tab:opcode-power-weight} Relative Opcode Power Consumption Weightings at 2.4 GHz }
\begin{tabular}{|c|c|}
\hline
\textbf{Category} & \textbf{Weight} \\
\hline
Addition & 1.43 \\
\hline
Multiplication & 1.74 \\
\hline
Division & 1.95 \\
\hline
Memory Read & 1.00 \\
\hline
Memory Write & 1.12 \\
\hline
Memory Copy & 2.84 \\
\hline
Comparison & 1.17 \\
\hline
Malloc & 21.06 \\
\hline
\end{tabular}
\end{table}

Below is an example of analyzing a program using our methodology. 
The C code for a scalar product of two arrays

\lstset{
  language=C,                                 
  stepnumber=1,                          
  numbersep=5pt,                  
  backgroundcolor=\color{white},  
  showspaces=false,               
  showstringspaces=false,         
  showtabs=false,                 
  tabsize=4,                      
  captionpos=b,                   
  breaklines=true,                
  breakatwhitespace=true,        
}
{\footnotesize
\begin{lstlisting}
double scalar_product(double A[], double B[], int n)
{  double sum = 0.0;
   for (int i = 0; i < n; i++) sum += A[i] * B[i];
   return sum;
}

\end{lstlisting}
\normalsize{
produces 
the following compiler optimized assembly code.}}
{\footnotesize
\lstset{
  language={[x86masm]Assembler},                                 
  stepnumber=1,                          
  numbersep=5pt,                  
  backgroundcolor=\color{white},  
  showspaces=false,               
  showstringspaces=false,         
  showtabs=false,                 
  tabsize=4,                      
  captionpos=b,                   
  breaklines=true,                
  breakatwhitespace=true,        
}

\begin{lstlisting}
scalar_product:
	testl	%r8d, %r8d           ; C1
	jle	.L4						 ; C7
	leal	-1(%r8), %r9d        ; C6
	movl	$0, %eax             ; C6
	pxor	%xmm0, %xmm0         ; C1
	jmp	.L3                      ; C7
.L5:
	movq	%r8, %rax            ; C6
.L3:
	movsd	(%rcx,%rax,8), %xmm1 ; C6
	mulsd	(%rdx,%rax,8), %xmm1 ; C3
	addsd	%xmm1, %xmm0         ; C1
	leaq	1(%rax), %r8         ; C6
	cmpq	%r9, %rax            ; C1
	jne	.L5                      ; C7
.L1:
	ret                          ; C0
.L4:
	pxor	%xmm0, %xmm0         ; C1
	jmp	.L1                      ; C7
\end{lstlisting}
}
\noindent
We identify five Category 1 instructions, one Category 3 instruction, 
five Category 6 instructions, and four Category 7 instructions,
which yields the weighted average instruction cost estimate
${5}/{15} + (1.03){1}/{15} + {5}/{15} + {4}/{15} = 1.002.$
In this example, mostly low cost operations are performed resulting in the expected instruction 
cost of $\sim$1.002. This standalone result is meaningless, but when juxtaposed with other 
measurements from other processes, relative instruction time disparities can be predicted and 
categorizations can be assigned. Likewise, the same procedure is utilized to estimate the cost of 
instructional power consumption.

\subsection{Preference List Generation} \label{preference}

Generation of preference lists 
for our bipartite matching 
is based on the two weight functions given below. 
First, combined weightings of the parameters for each $u \in U$ and each $v \in V$ must be 
created. As defined in Section~\ref{parameters}, the parameters of these 
sets are stored in $P$ and $C$ respectively. 
\begin{definition}[Task Weight Function]
The weighting $w_p$ for each $p\in P$ is defined as
$ w_p (p) = {1}/{2}\left({p_1}/{3} + {p_2}/{3}\right).$
\end{definition}

\begin{definition}[Core Weight Function]\label{core-weight-function}
The weighting $w_c$ for each $c\in C$ is defined as
\begin{align*}
& w_c (c, C) \\
& \quad = \frac{1}{3} \left(\frac{c_1}{\max\{c_1\}}
 + \left(1 - \frac{c_2}{\max\{c_2\}} \right) +  \left( 1-\frac{c_3}{\max\{c_3\}} \right) \right)
\end{align*}
where each maximum is taken over all $c\in C$.
\end{definition}

In generating preference lists, the favorability of a task-core match is defined as 
\textit{matching affinity}.
Matching affinity value is a real number in the range $[0,1]$ with $1$ being the highest affinity 
and $0$ the lowest. There are two forms of matching affinity.
The first matching affinity is from a task to a core, for which function $w_c$ in 
Definition~\ref{core-weight-function} is used.
\begin{proposition}[Matching Affinity from Task to Core]
Function $w_c$ represents the matching affinity for the tasks to each processor core.
\begin{proof}
Tasks themselves greedily develop their preference lists
in a way that each task wishes to choose the most adept processor core available in which to 
execute. Therefore, the preference list of each task during an accumulation window 
is identical.
By construction, the function $w_c$ in Definition~\ref{core-weight-function} results 
in the highest value for the core with the lowest net stress, 
indicating that it is preferred the most.
Therefore, function $w_c$ represents the matching affinity for the task to each processor core.
\end{proof}
\end{proposition}

The second matching affinity is from a core 
to a task 
that quantifies the preference for a particular task from the perspective of a core. 

\begin{definition}[Matching Affinity from Core to Task] \label{def:matching-core-task}
The matching affinity from a core with parameters $c \in C$ to a task with parameters $p \in P$ is defined as
\begin{align*}
m(p,c,P,C) &= \beta \left(1 - \frac{| w_p (p) - w_c (c, C) |}{\max{\{w_p (p), w_c (c, C)}\}}\right)\\
&+ \frac{1 - \beta}{2} \left(\left( 1- \frac{p_3}{\max\{p_3\}}\right) + \frac{p_c}{\max\{p_c\}}\right) 
\end{align*}
where $\max\{p_3\}$ and $\max\{p_c\}$ are 
each taken over all $p\in P$.
The constant $\beta$ 
takes a value within $[0, 1]$ to take a weighted average of 
the core-centric matching scheme (similarity between core and task weight functions) 
and the task-centric metrics of credit and deadline. By default,
$\beta = 1/2$  
to equally weight both components of $m$. Values of $\beta < 1/2$  
increase the influence of deadline fulfillment and credit
fairness in the matching, potentially at the expense of equitable core 
performance. On the contrary, values of $\beta > 1/2$   
emphasize 
matching based on core load and temperature balancing, potentially leading to 
better system performance but 
an increased rate of missed deadlines and delay in task execution.
\end{definition}

\begin{theorem}[Optimality of Core to Task Matching Affinity] \label{optimality-task-matching-affinity}

The function $m$ described in Definition~\ref{def:matching-core-task}
provides an optimal metric to generate core preference lists for usage
in the task-core matching based on a modified Gale-Shapely algorithm.

\begin{proof}
The right-hand side of the matching affinity equation for $m$ 
can be subdivided into two parts, affinity based on
the core and task weight functions
\begin{equation}
1 - \frac{| w_p (p) - w_c (c, C) |}{\max{\{w_p (p), w_c (c, C)}\}} \label{part1}
\end{equation}
and affinity based on task-centric credit and deadline
\begin{equation}
\frac{1}{2} \left(\left( 1- \frac{p_3}{\max\{p_3\}}\right) + \frac{p_c}{\max\{p_c\}}\right). \label{part2} 
\end{equation}
A weighted average of~(\ref{part1}) and~(\ref{part2})
are taken with a constant $\beta$ to define the full matching affinity $m$. 

In~(\ref{part1}),
$ {| w_p (p) - w_c (c, C) |}/{\max{\{w_p (p), w_c (c, C)}\}}$
uses the mathematical concept of relative difference, whereby the magnitude
of the difference between core and task weight functions are divided by 
the maximum between the two. By 
definition of this mathematical comparative measurement and the 
fact that both weight 
functions return a nonnegative value, the result gives a normalized
difference in the range $[0, 1]$ where $0$ indicates absolute similarity and $1$ 
suggests maximal difference (one of the two numbers is $0$).
Affinity measures similarity, the inverse of difference, so the resultant
value is subtracted from $1$ yielding~(\ref{part1}).

In~(\ref{part2}) with regards to deadline, a 
ratio of the current deadline to the maximum is added to the affinity. 
Earlier deadlines incur higher priority, so  
in order to weight the deadline within the range $[0, 1]$ with $1$ 
being highest priority, we take
$1 - ({p_3}/{\max{\{p_3\}}}).$
With regards to task credit, we proportionally take the current task's credit relative 
to the maximum: 
$ {p_c}/{\max{\{p_c\}}}.$
Then, the average of these two terms is taken to yield~(\ref{part2}).

The $\beta$ term 
expresses the desired weighting of the core-centered 
metrics and task-centered metrics.
Its default value of $1/2$
means that the resulting matching affinity
is a simple average of all the weight metrics utilized. 
 
Once all values of function $m$ are calculated for all $c\in C$ 
and for all $p\in P$, the values
for each core are sorted in a descending order, resulting in
the core's preference list, 
which is used in the modified Gale-Shapely algorithm (described in Section~\ref{matching}) 
to produce an optimal task-core matching.   
\end{proof}
\end{theorem}

\subsection{Matching Algorithm} \label{matching}

We implement a bipartite matching algorithm based upon the Gale-Shapely algorithm~\cite{1989827}. The original Gale-Shapely algorithm 
produces a stable marriage matching between a group of men and a group of women where the two groups are of the same size. Each individual has a list of preferences ranking the other group.
A marriage matching is then assigned between men and women based on these preferences. A matching is considered stable if there exists no man and woman pair such that they mutually prefer each other over their current partner.

Our algorithm differs from Gale and Shapely's initial algorithm in 
the sense that
a matching is performed on a set $U$ of process tasks and a set $V$ of available processor cores where the size of the two sets, $|U|$ and $|V|$, may or may not be equal. Additionally, our algorithm incorporates dynamic data as part of preference data.
Based upon the categorizations provided by the program analysis, available processor cores provide a preference listing based on their current state as defined in the parameters of interest for the processor. 
The schema for generating preference lists is described in Section~\ref{preference} and promotes cooperation among processor cores. 
After we have obtained the preference lists, each task is stored in a queue. Each task then proposes to its highest ranked core which it has not yet proposed to. If the core accepts the proposal, then the task is dequeued. 
If the core has a current partner 
but prefers the new proposal, then it is unpaired with its current partner (the current partner becomes an old partner), paired with the new partner, and the old partner returns to the queue. The process is repeated until either all tasks are matched, in the case of $|U|\le |V|$, or all valid proposals have been made, in the case of $|U|>|V|$.
The modified Gale-Shapely algorithm is given in Algorithm~\ref{alg:task-core-matching}.

For an example of matching between tasks and processor cores, consider tasks 1--5 to be scheduled to processor cores 1--4 in which no task specifies an early deadline. Suppose we have already generated preference lists for each processor core and task as in Table~\ref{tab:T5}.
\vspace*{-5pt}
\begin{table}[h]
\centering
\caption{\label{tab:T5} Bipartite Matching Preference}
\vspace*{-5pt}
\begin{tabular}{|c|c|c|}
\hline
\textbf{ID} & \textbf{Task Preference} & 
\textbf{Core Preference}\\
\hline
1 & (2, 1, 3, 4) & (3, 1, 4, 2, 5) \\
\hline
2 & (2, 1, 3, 4) & (2, 4, 3, 1, 5) \\
\hline
3 & (2, 1, 3, 4) & (3, 2, 1, 4, 5) \\
\hline
4 & (2, 1, 3, 4) & (1, 4, 3, 5, 2)\\
\hline
5 & (2, 1, 3, 4) & NONE \\
\hline
\end{tabular}
\end{table}
Then, matching is performed using 
Algorithm~\ref{alg:task-core-matching} as shown in Table~\ref{tab:T6}.
\vspace*{-5pt}
\begin{table}[H]
\centering
\caption{\label{tab:T6} Bipartite Matching Evaluation }
\vspace*{-5pt}
\begin{tabular}{|c|l|}
\hline
\textbf{Iteration} & \textbf{Proposals} \\
\hline
$1$ & $(1 \rightarrow 2), \mathbf{[2 \rightarrow 2]}, (3 \rightarrow 2), (4 \rightarrow 2), (5 \rightarrow 2)$ \\
$2$ & $(1 \rightarrow 1), \mathbf{[3 \rightarrow 1]}, (4 \rightarrow 1), (5 \rightarrow 1) $ \\
$3$ & $\mathbf{[1 \rightarrow 3]}, (4 \rightarrow 3), (5 \rightarrow 3)$ \\
$4$ & $\mathbf{[4 \rightarrow 4]}, (5 \rightarrow 4)$ \\
\hline
\end{tabular}
\end{table}
\noindent The following bipartite graph represents the matching.
\vspace*{-5pt}
\begin{figure}[H]
\centering
\resizebox{0.4\textwidth}{!}{%
\begin{tikzpicture}[node distance=1mm and 35mm, main/.style = {draw, circle}]
\node[main](1) {$1$};
\node[text width=4.5cm] at (0, -1) {\hspace{7mm} Tasks};
\node[main](2) [below=of 1] {$2$};
\node[main](3) [below=of 2] {$3$};
\node[main](4) [below=of 3] {$4$};
\node[main](5) [below=of 4] {$5$};

\node[text width=2.5cm] at (6,-1) {Cores};
\node[main](6) [right=of 1] {$1$};
\node[main](7) [right=of 2] {$2$};
\node[main](8) [right=of 3] {$3$};
\node[main](9) [right=of 4] {$4$};

\draw[-] (3) -- (6);
\draw[-] (2) -- (7);
\draw[-] (1) -- (8);
\draw[-] (4) -- (9);

\end{tikzpicture}
}%
\end{figure}
\vspace*{-5pt}

\begin{algorithm}
\begin{algorithmic}[1]
\caption{Modified Gale-Shapely algorithm for matching}
\label{alg:task-core-matching}
\State{\textbf{procedure} Task-Core-Matching{$(U, V, L_u, L_v)$}}
\Statex{$\triangleright$ $U$: set of tasks to be scheduled, $V$: set of available processor cores, $L_u$ and $L_v$: preference lists of $U$ and $V$}
\Statex{$\triangleright$ $\gets$ operator: variable assignment} 
\State{$E \gets \text{empty list of edges in matching}$}
\For{each $v \in V$} \Comment {Initialization}
	\State{$v.partner \gets NIL$}
\EndFor

\For {each $u \in U$}
	\State{$u.partner \gets NIL;$\, $u.proposed[\forall v \in V] \gets$ False}
\EndFor

\While {$\exists u_{\in U} (u.partner \text{ is } NIL)\land \exists v_{\in V} (\neg u.proposed[v])$}
	\State{$v \gets \text{first unproposed } v \in L_u [u]$}
	\State{$u.proposed[v] \gets $ True}
	\If{$v.partner \text{ is } NIL$}
		\State{$u.partner \gets v;$\, $v.partner \gets u$}
	\Else
		\State{$u^\prime \gets v.partner$} \Comment{$u^\prime$ is $v$'s current partner}
		\If{$L_v[v].\text{indexOf}[u] < L_v[v].\text{indexOf}[u^\prime]$}
			\State{$u.partner \gets v;$\,  $v.partner \gets u$}
			\State{$u^\prime.partner \gets NIL$}
		\EndIf
	\EndIf
\EndWhile

\For{each $u \in U$}
	\If{$u.partner \text{ is not } NIL$}
		\State{$E \gets E \cup \{(u, u.partner)\}$}
	\EndIf
\EndFor
\State\Return{$E$}

\end{algorithmic}
\end{algorithm}
\vspace*{-10pt}
\subsection{Dynamic Windowing} \label{dynamic}
A self-adjusting dynamic window of accumulation time 
is used to collect processes ready to be scheduled. As processes arrive, they are required to wait as other processes are collected. Once the window of accumulation time has 
elapsed, the matching is performed and processes are shipped for execution in a time period known as the execution window. Immediately following the end of the first accumulation window, a new accumulation window of the same size begins. The process continues until the end of the first execution window in which 
resizing occurs based on 
the performance of the first execution window along with previously finished executions of the same accumulation window size.
These results are compared to the execution performance of the previous accumulation window size. Based on this comparison, the window size will either increase, decrease, or remain the same for the next accumulation window. Initially, as tasks are scheduled early within our program, fluctuations of window size will be relatively large as the optimal nature of windowed scheduling is not yet known to the system. As  
dynamic window scheduling continues and predictable behavior in process arrival persists, the change in window size between schedules will decrease. The window will eventually be able to rely almost entirely on previous window size and not change with continuation of the scheduling trend. However, if a significant change occurs to process arrival resulting in decreases in matching satisfaction (described in Section~\ref{performance}), the window size will begin to change again.

Optimization of processor 
utilization in completing tasks while ensuring a notion of fairness and equitability is the ultimate goal of the scheduling. 
Under high strain and pressure, a CPU 
begins throttling processes and activates self-protection mechanisms which 
lower the rate of a process' completion~\cite{7000001}. 
Our algorithm ensures load balancing to account for CPU core temperature spikes and manage operating temperature. 
We additionally consider process waiting time via the credit 
system to ensure fairness. We first define two throughput metrics 
that drive the resizing process. 

\begin{definition}[Utilization Throughput] \label{utilization}
The \textit{utilization throughput} is defined as the average holistic core utilization throughout 
an execution time window and is calculated as 
\begin{center}
$ UT (C) = (\sum_{c \in C} c_2)\,/\,|C|$
\end{center}
where the value of $UT$ is in the range $[0,1]$.
\end{definition}
Within an execution time window, the average throughput of 
each core is monitored and stored in its $c_2$. Inter-core differences in terms of throughput 
are considered in 
task-core matching. However, when considering system throughput, we must take an average of the throughput of all cores in the system for the current 
execution window. 

\begin{definition}[Operating Throughput] \label{operating}
The \textit{operating throughput} is defined as the 
average speed of a 
core relative to its maximum potential during an execution time window 
\begin{center}
$ OT (C) = \left(\sum_{c \in C} ({c_1}/{c_{msp}})\right)/\, |C|$
\end{center}
where the value of $OT$ is in the range $[0,1]$.
\end{definition}
High operating throughput implies that the processor is running at a high clock speed. 
This measurement is included mostly to prevent processor 
overheating in extreme instances. For instance, our system would not wish to run processor cores at $100\%$ 
utilization if the CPU is being throttled to a significantly lower speed due to excessive CPU temperature. In such a scenario, we would wish to reduce burden on our processor
by increasing accumulation window size to allow for a cool-down period and appropriately manage scheduling among cores to prevent throttling. 

A processor may alter its speed in two 
scenarios. The first case is when excessive usage occurs. If a processor is pushed too hard and exceeds temperature limits, low-level CPU software will throttle individual or all processor core speeds in order to attempt to lower the processor temperature. In drastic scenarios, 
the operating system will order a CPU core to cease all activity in a self-preservation attempt. The second case in which a processor core will reduce speed is when there is a high idle rate among the processor cores. In such a scenario, there is no need to run processor cores at full capacity while only performing simple tasks, so processors reduce speed to save energy. 

In the first case, our system will increase the accumulation window size to assist in the throttling process by allowing the processor to cool-down. In the second case of reduced processor speed, we will reduce the accumulation time window to increase responsiveness when a task does arrive. It is expected that in most cases of normal processor functioning when executing tasks, operating throughput will be maximized yielding a measurement of $1$. 

\subsubsection{Window Performance and Matching Performance} \label{performance}

We define a \textit{window performance} metric to assess the 
CPU throughput and the ability of tasks to be scheduled at the end of an accumulation window. 
This metric is used to determine when a matching is shipped for processor execution within our system. 
Window performance is contingent on the two measurements $UT$ and $OT$ defined in Definitions~\ref{utilization} and~\ref{operating}. Each of these measurements is weighted with a constant such that the 
resulting aggregate measurement is in the range $[0,1]$. A measurement of $1$ for window performance is indicative of a perfect execution window. This scenario would occur when during an execution window, $100\%$ of all CPU cores are utilized, all cores are open to scheduling at the end of an execution window, and all cores are operating at maximum speed with adequate temperature control. 

\begin{definition}[Window Performance] \label{window-perf-metric}
The total aggregate weighting given to the performance of an
execution window corresponding to a particular accumulation window size with respect to processor 
cores 
is defined as the \textit{window performance} ($WP$) which is in the range $[0, 1]$:
$$ WP (C) = \alpha_2 UT (C) + \alpha_1 OT (C) $$
where $\alpha_1$ and $\alpha_2$ are positive constants whose sum is $1$. 
The average performance of the first execution window along with execution windows that finish earlier than the first and correspond to the same accumulation window size is defined as $AWP$. 
\end{definition}

\begin{proposition}[Optimal Window Performance Constants] \label{equal-constants}
The optimal constants 
for $WP$ in Definition~\ref{window-perf-metric} are 
$\alpha_1 = \alpha_2 = 1/2$.
$\alpha = (1-\alpha) = 1/2$.
\begin{proof} 
Substituting 
the definitions of utilization throughput and operating throughput (Definitions~\ref{utilization} and~\ref{operating}) into the definition of window performance (Definition~\ref{window-perf-metric}) yields
$
WP(C) = \alpha_2 (\sum_{c \in C} c_2)\,/\,|C| + \alpha_1 \left(\sum_{c \in C} ({c_1}/{c_{msp}})\right)/\, |C|
$
where
$\alpha_1 + \alpha_2 = 1$ and
$\forall {c\in C} [c_1, c_{msp} \in (0, \infty ), c_{msp} \geq c_1, \text{ and } c_2 \in [0,1]]$.
Rearranging the above equation yields
$WP (C) = (\sum_{c \in C} (\alpha_2 c_2 + \alpha_1  {c_1}/{c_{msp}}))\,/\, |C|,$
and letting 
$u = {c_1}/{c_{msp}} \text{ for each } c\in C$ and substituting above
yields 
\begin{center}
$WP (C) = (\sum_{c\in C} (\alpha_2 c_2 + \alpha_1 u))\,/\, |C|$ \label{eqn:wp}
\end{center}
where
$\alpha_1 + \alpha_2 = 1$ and 
$\forall c\in C ( c_2, u \in [0,1] )$.
The final equation formulates the window performance calculation as a linear combination 
of variables $c_2$ and $u$, each value in the range of $[0,1]$.
It follows that the optimal throughput measure is obtained by
having an equal weighting between $c_2$ and $u$,
i.e., $\alpha_1 = \alpha_2 = 1/2$,
maintaining equal weight between the utilization throughput and the operating throughput.
\end{proof}
\end{proposition}

An additional metric utilized and equally weighted by the window resizing procedure is the \textit{matching performance} relative to the tasks. This performance 
measure is defined as the average waiting time of the tasks until being scheduled. 
Once scheduled, tasks 
are run to completion without preemption.
\begin{definition}[Matching Performance] \label{matching-performance}
For a matching $M$ executed at time $t$, the matching performance is defined as
\begin{center}
$ MP (M, t) = (\sum_{(u, v) \in M} (t - u_{art}))\,/\, |M| $
\end{center}
where $M$ is the set of 
$(u,v)$ pairs, each of which represents task $u$ being 
scheduled to run on processor $v$, $ u_{art} $ is the arrival time of 
$u$, and $t$ is the current time in which the matching is executed. 

The average performance of the first execution window along with subsequent execution windows that finish earlier and correspond to the same accumulation window size is defined as $AMP$.
\end{definition}

When performing matchings within an accumulation time window there are several notions which should be clarified. Firstly, 
there are the start and end times of a time window. The time window resizing solely defines the length of the window and not the start and end times. The next 
accumulation window 
always begins upon the first task arrival after the current time period ends. If the previous time window ends at time $t_1$, then the next window will begin at $t_1^\prime > t_1$. If there already exist tasks in the waiting queue at time $t_1$, then the next window will begin at $t_1^\prime = t_1 + \epsilon$ where $\epsilon$ represents the small finite time period in which the time window resize calculation is performed. Secondly, the time utilized to determine 
window performance is the total time elapsed from the previous window's end time to the current window's end time.

\begin{proposition}[Window Resizing Direction] \label{window-resizing-dir}
The following three cases cover the different
scenarios of window resizing.

\smallskip

\noindent
\underline{Window Resizing Case 1:} \label{window1}
If $\frac{1}{2}\left(\frac{AWP_{cur}}{AWP_{prv}} + \frac{AMP_{prv}}{AMP_{cur}}\right) > 1$, then the window resize proved beneficial to the system. 
Thus, continue resizing in the same direction as before for the next accumulation window.

\smallskip

\noindent
\underline{Window Resizing Case 2:} \label{window2}
If $\frac{1}{2}\left(\frac{AWP_{cur}}{AWP_{prv}} + \frac{AMP_{prv}}{AMP_{cur}}\right) = 1$, then the current window is operating at the same performance level as the previous. In this scenario, no resizing is necessary.

\smallskip

\noindent
\underline{Window Resizing Case 3:} \label{window3}
If $\frac{1}{2}\left(\frac{AWP_{cur}}{AWP_{prv}} + \frac{AMP_{prv}}{AMP_{cur}}\right) < 1$, then the previous time window performed better than the current, indicating that we must reverse the direction in which the window resizes for the next accumulation window. 

\begin{proof}
Substituting 
the definitions of $UT$ and $OT$ into the formula for $WP$ in Definition~\ref{window-perf-metric} and using the optimal constants $\alpha_1 = \alpha_2 = 1/2$ shown in Proposition~\ref{equal-constants} yields\\
$WP(C) = \left(\sum_{c \in C} c_2\right)/ 2 |C| +  \left(\sum_{c \in C} ({c_1}/{c_{msp}})\right)/ 2 |C|$
which yields
$\left(\sum_{c \in C} (c_2 + ({c_1}/{c_{msp}} ))\right)/ 2 |C|$
by rearranging terms.

Now, consider the ratio of window performance measurements for two 
unique windows,  
\begin{align*} 
R_{wp} = \frac{WP (C_1)}{WP (C_2)} &= \frac{[\sum_{c \in C_1} (c_2 + ({c_1}/{c_{msp}} ))]\,/\, 2 |C_1|}
{[\sum_{c \in C_2} (c_2 + ({c_1}/{c_{msp}} ))]\,/\, 2 |C_2|}\\
&= \frac{[\sum_{c \in C_1} (c_2 + ({c_1}/{c_{msp}} ))]\,/\, |C_1|} {[\sum_{c \in C_2} (c_2 + ({c_1}/{c_{msp}} ))]\,/\, |C_2|}
\end{align*}
where the two sets $C_1$ and $C_2$ are sets at points $t_1$ and $t_2$ in time, respectively.

The ratio $R_{wp}$ between two window performances reflects 
the ratio between the overall averages of equally weighted combinations of CPU core utilization $c_2$ and CPU core speed $c_1$ relative to the maximum.
The ratio 
$R_{wp} = 1 $ means that the 
overall performance of the windows are equal. 
The ratio 
$R_{wp} < 1$ indicates that $WP(C_2)$ performed 
better than $WP(C_1)$
and 
$R_{wp} > 1$ means that 
$WP(C_1)$ performed better than $WP(C_2)$.

Now, consider matching performance 
defined in Definition~\ref{matching-performance}
and take the ratio of two unique matching performance measurements
\begin{equation*}
R_{mp} = \frac{MP (M_1, t_1)}{MP (M_2, t_2)} = \frac{[\sum_{(u_1, v_1) \in M_1} \left( t_1 - 
u_{1,art} \right)]\,/\,|M_1|}{[\sum_{(u_2, v_2) \in M_2} \left( t_2 - u_{2,art} \right)]\,/\,|M_2|}. \label{eqn:mp-ratio}
\end{equation*}
Matching performance is measured based on task's waiting time, which we wish to minimize.
Thus, the matching performance measurement with the minimal result is regarded as performing the best. Therefore, 
$R_{mp} = 1$ means that matching performance between the two measurements is identical, 
that is, the average waiting times were equal. If 
$R_{mp} < 1$, then $MP (M_1, t_1)$ performed the best as its average waiting time 
is lower than $MP (M_2, 
t_2)$ and if 
$R_{mp} > 1$, then $MP (M_2, t_2)$ performed the best.

To equally weight contributions from both window and matching performances, the 
measurements of $AWP$ and $AMP$ are averaged together. 
Based on the resultant value of this average, we greedily determine
the next window size based on performance of the previous and the current 
direction of resizing. 
It follows based on the analysis of window and matching 
performance metrics that the provided formulation satisfies our greedy resizing 
scheme. 
\end{proof}
\end{proposition}

Utilizing the direction of window 
resizing 
based on the ratio of performance benefit derived in Proposition~\ref{window-resizing-dir}, 
we scale up or down the
accumulation window size in 
an attempt to reach steady state convergence of the time window 
around an ideal sizing. 
We achieve this behavior with our next proposition by providing
the formulae used to obtain  
a more desirable next window size.

\begin{proposition}[Next Window Size Heuristic]
The next window size is calculated as a scaled factor of the current window size 
relative to the previous size, defined by the two cases below.

\smallskip

\noindent
\underline{Next Window Size Case 1:}\,\, \label{next1}
If $\Delta t_{cur} \geq \Delta t_{prv}$, then
\begin{center}
$ \Delta t_{nxt} = \frac{\Delta t_{cur}}{2} \left(\frac{AWP_{cur}}{AWP_{prv}} + \frac{AMP_{prv}}{AMP_{cur}}\right). $
\end{center}
\noindent
\underline{Next Window Size Case 2:}\,\, \label{next2}
If $\Delta t_{cur} < \Delta t_{prv}$, then
\begin{center}
$ \Delta t_{nxt} = \frac{\Delta t_{cur}}{2} \left(\frac{AWP_{prv}}{AWP_{cur}} + \frac{AMP_{cur}}{AMP_{prv}}\right). $
\end{center}
\begin{proof}
We first consider Next Window Size Case $1$. 
As observed in Proposition~\ref{window-resizing-dir}, 
the expression
$ R = ({1}/{2}) \left( {AWP_{cur}}/{AWP_{prv}} + {AMP_{prv}}/{AMP_{cur}} \right)
$
gives the ratio of improvement or degradation of performance between $\Delta t_{prv}$ and $\Delta 
t_{cur}$. It 
follows that in the occurrence of performance benefit, the provided 
formula increases the window size in proportion to the measured benefit as 
$R  > 1$
and $\Delta t_{cur} \geq \Delta t_{prv}$ 
indicates that the previous direction 
of resizing was increasing. 
In the case of performance degradation, the provided formula correctly
decreases the window size in accordance with Proposition~\ref{window-resizing-dir}
as $R < 1$.

Now, consider Case $2$ in which $\Delta t_{cur} < \Delta t_{prv}$. In this instance,
performance benefit indicates that continuing to reduce the window size yields 
benefit and performance degradation suggests that 
increasing the window size is 
necessary based upon Proposition~\ref{window-resizing-dir}. Therefore, to 
achieve the opposite resizing effect from Case $1$, it is necessary 
to take the inverse of the utilized ratios to properly resize in the correct
direction. It follows that the appropriate window resizing scheme can be given
by the following formula: 
\begin{center}
$\Delta t_{nxt} = \frac{\Delta t_{cur}}{2} \left(\left(\frac{AWP_{cur}}{AWP_{prv}}
\right) ^{-1} + \left( \frac{AMP_{prv}}{AMP_{cur}}\right) ^ {-1} \right)$\\
$= \frac{\Delta t_{cur}}{2} \left(\frac{AWP_{prv}}{AWP_{cur}} + \frac{AMP_{cur}}
{AMP_{prv}}\right).\phantom{\frac{A_{ci}}{A_{pi}}}$ 
\end{center}
\vspace*{-7pt}
\end{proof}
\end{proposition}
We provide 
the algorithm for the window resizing subroutine 
in Algorithm~\ref{resize-alg}.
The previous matching, current matching, previous time window size, and current window size are taken as parameters by the procedure. The algorithm calculates the average window performance  
using Definition~\ref{window-perf-metric} and average waiting time 
using Definition~\ref{matching-performance}, then produces the next time window size based on the cases described in Proposition~\ref{next1}.

\subsubsection{Execution Window Cutoff} \label{cutoff}

In most circumstances of our algorithm, the execution window 
ends when all of the scheduled tasks from the matching are completed. This windowing scheme works well for most cases but struggles when an execution encounters a long-running task that is an upper outlier compared to other scheduled tasks in terms of execution time. 
In order to prevent such 
situations where the next window resizing 
is indefinitely prolonged while a single task nonrepresentative of previous schedulings fails to end,
we establish a cutoff for the tasks. We utilize Tukey's upper, outer fence for \textit{far out} values as a cutoff for the execution window. Based on recordings of previously scheduled tasks, the execution window will end after $Q_3 + 3 (Q_3 - Q_1)$ seconds have elapsed where $Q_3$ is the third quartile of previous execution times, and $Q_1$ is the first quartile of previous execution times. This formulation of outlier identification is a commonly used scheme first identified in~\cite{exploratory-dta-analysis}.

\begin{algorithm}[H]
\begin{algorithmic}[1]
\caption{Window Resizing Procedure}
\label{resize-alg}
\State{\textbf{procedure} Window-Resize{$(M_{prv}$, $M_{cur}$, $\Delta t_{prv}$, $\Delta t_{cur})$}}
\Comment {$M$ is matching, $\Delta t$ is window size}
\State{$\Delta t_{nxt} \gets NIL$} \Comment{$\Delta t_{nxt} $ is the next window size}
\State{$AWP_{prv} \gets \text{Average-Window-Performance} (M_{prv})$}
\State{$AWP_{cur} \gets \text{Average-Window-Performance} (M_{cur})$} 

\State{$AMP_{prv} \gets \text{Average-Waiting-Time} (M_{prv})$}
\State{$AMP_{cur} \gets \text{Average-Waiting-Time} (M_{cur})$}

\If{$\Delta t_{cur} \geq \Delta t_{prev}$} \Comment{Proposition~\ref{next1} Case 1}
	\State{$\Delta t_{nxt}  \gets \frac{\Delta t_{cur}}{2}\left(\frac{AWP_{cur}}{AWP_{prv}} + \frac{AMP_{prv}}{AMP_{cur}}\right)$}

\Else \Comment{Proposition~\ref{next2} Case 2}
	\State{$\Delta t_{nxt}  \gets \frac{\Delta t_{cur}}{2}\left(\frac{AWP_{prv}}{AWP_{cur}} + \frac{AMP_{cur}}{AMP_{prv}}\right)$}
	
\EndIf

\State\Return{$\Delta t_{nxt} $}

\end{algorithmic}
\end{algorithm}

\subsubsection{Window Resizing Example} \label{example}
\begin{figure}
\centering
\includegraphics[scale=.78]{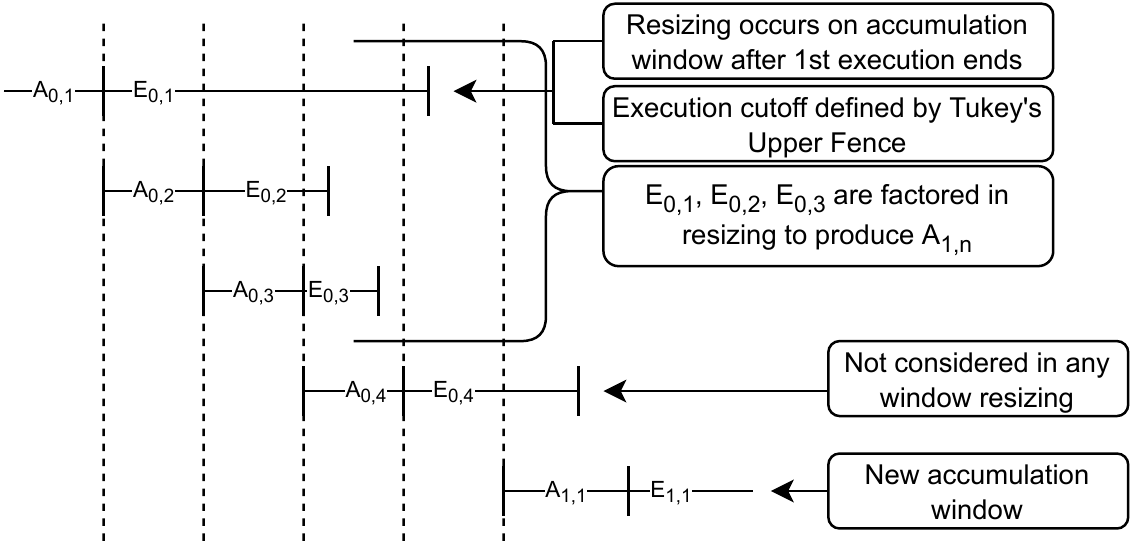}
\caption{Window resizing example.}
\label{fig:window-resizing-diagram}
\end{figure}

The following example is illustrated in the diagram in Fig.~\ref{fig:window-resizing-diagram}.
As described in Section~\ref{time}, our scheduler consists of accumulation windows followed by execution windows. Accumulation windows are periods of waiting time in which the scheduler waits for tasks arrive to make an informed decision on which to schedule. Suppose that we have an accumulation window $A_{0,1}$. Our scheduler waits until the accumulation window ends and then performs an execution $E_{0,1}$. Accumulation window resizing 
occurs when $E_{0,1}$ ends by either all tasks finishing execution or the cutoff time limit being reached. However, to prevent excessive idle time from freed cores, a new accumulation window $A_{0,2}$ of the same size starts immediately after $A_{0,1}$ ends and $E_{0,1}$ begins. The algorithm performs this way to ensure that an accumulation window is always occurring during runtime of the scheduler (excluding the time during preference list generation and matching). 
In this instance, four accumulation windows of the same size ($A_{0,1}$, $A_{0,2}$, $A_{0,3}$, and $A_{0,4}$) are deployed. When execution window $E_{0,1}$ ends, we see that $E_{0,2}$ and $E_{0,3}$ also have ended. Therefore, when resizing the next accumulation window $A_{1,n}$ for $n \geq 1$, we 
utilize averaged results between $E_{0,1}$, $E_{0,2}$, and $E_{0,3}$ relative to execution windows corresponding to a previous accumulation window size (not shown in this diagram). Because $E_{0,4}$ ends after $E_{0,1}$ ends, it is not factored into any window resizing but is allowed to complete its execution window.

\subsubsection{Task Scheduler} \label{scheduler}

We now give the Task Scheduler subroutine in Algorithm~\ref{alg:schedule} which drives 
preference list generation, matching, and window resizing.
Initially, our scheduler performs an initialization of thread-safe arrival buffers for both executable tasks and processor cores. 
Secondly, initialization of the first accumulation window occurs. The initial time window consists of the time between the first and second task. This window size is assumed to be a low estimate by our system, so the next window size is always fixed as double the time between the first two arrivals. Defaulting to this behavior enables exponential growth of our time window by an initial power of $2$. This 
initial exponential growth of the time window size 
allows the window to be quickly scaled to an appropriate level which can then be fine-tuned as performance reaches the ideal, optimal time period based on current trends of arrival. The initialization is finished by computing preference lists for tasks and cores followed by the first execution. Once initialization is completed, our scheduler runs in 
an infinite while loop. Within this while loop, a nested while loops runs while the initial execution of the first instance of the current accumulation window is running. The appropriate window time period is waited, followed by computation of preference lists, matching, and execution of the matching. Matchings that occur after the first instance of a given accumulation window size are referred to as
$M_{mid}$.
After the first execution of the current accumulation window completes, the window is resized based upon measured performance metrics described in Definitions~\ref{window-perf-metric} and~\ref{matching-performance}, and Proposition~\ref{next1}. The new window then begins, followed by a matching, and execution. The loop then iterates and the process repeats.
The Task-Core-Matching subroutine used in lines~15 and 23 is from Algorithm~\ref{alg:task-core-matching} and the Window-Resize subroutine in line~19 is from Algorithm~\ref{resize-alg}.

\vspace*{-3pt}
\begin{algorithm}
\begin{algorithmic}[1]
\caption{Scheduling Algorithm}
\label{alg:schedule}
\State{\textbf{procedure} Task-Scheduler$()$}
\State{Initialize unbounded-buffer $U$ for task arrival}
\State{Initialize bounded-buffer $V$ for available cores}

\Statex{$\triangleright$ Initialization of Accumulation Window}
\State{$t_0 \gets$ time of first task arrival}
\State{$t_1 \gets$ time of second task arrival}
\State{$M_{prv}, M_{cur} \gets$ matching of first task arrivals}
\State{$\Delta t_{prv} \gets t_1 - t_0$}
\State{$\Delta t_{cur} \gets 2 \Delta t_{prv}$}

\State{Compute $L_u$ and $L_v$} \Comment{First Execution Window}
\State{EXECUTE $M_{cur}$}

\While{True}
	\While{$M_{cur}$ is still executing}
		\State{Wait until $\Delta t_{cur}$
		has elapsed since first arrival} 
		\State{Compute $L_u$ and $L_v$}
		\State{$M_{mid} \gets$ Task-Core-Matching$(U, V, L_u, L_v)$}
		\State{EXECUTE $M_{mid}$}
	\EndWhile
	
	\State{$\Delta t_{prv} \gets \Delta t_{cur}$}
	\State{$\Delta t_{cur}\gets$ Window-Resize$(M_{prv}, M_{cur}, \Delta t_{prv}, \Delta t_{cur})$} 
	\State{Wait until $\Delta t_{cur}$
	has elapsed after first task arrival}
	\State{Compute $L_u$ and $L_v$}
	\State{$M_{prv} \gets M_{cur}$}
	\State{$M_{cur} \gets$ Task-Core-Matching$(U, V, L_u, L_v)$}
	\State{EXECUTE $M_{cur}$}

\EndWhile

\end{algorithmic}
\end{algorithm}

\section{Experiments} \label{experiments}

In this section, we describe our testing framework, the control schedulers for comparison, the tasks implemented for scheduling, and explain the experimentation results.

\subsection{Non-preemptive Scheduling Algorithms Tested} \label{schedulings}
In order to test the performance of our algorithm relative to other scheduling algorithms, we implemented several alternate non-preemptive scheduling algorithms and tested with identical input data. All scheduling algorithms are implemented in C++20 with the GNU GCC compiler.

The first algorithm assessed is the First Come First Serve (FCFS) 
algorithm which is implemented through the creation of an unbounded, thread-safe buffer 
using a queue data structure. The buffer is shared across threads that each corresponds 
to a processor core. When a core is free for scheduling, the corresponding thread  pops the first-in-line task from the shared queue if possible and executes it. Otherwise, the thread waits for a new schedulable task. 

The second algorithm tested is a non-preemptive Earliest Deadline First (EDF) algorithm
which is implemented in a manner similar to FCFS with the exception that a priority queue implemented with a minimum heap is utilized in our unbounded buffer and priority is assigned based on a process' desired completion time. The desired completion time parameter is deemed optional and tasks that do not provide a completion time are given lowest priority.

The third algorithm tested is a non-preemptive Shortest Job First (SJF) algorithm. This algorithm makes use of our static assembly analyzer in order to classify tasks as requiring $low$, $moderate$, or $high$ instructional time categorization. Within these categorizations, execution occurs in a FCFS basis where the $low$ tasks are executed in order, followed by the $moderate$ tasks, and finally the $high$ tasks. This algorithm differs slightly from a traditional shortest job first in the sense that execution time lengths of processes are created by our system's categorization procedure for the average instruction cost.

The final scheduling algorithm tested is our Non-preemptive Dynamic Windowing (NPDW) described in Section~\ref{algorithm}. It is implemented following the pseudocode provided in this paper.

\subsection{Testing Methodology} \label{testing}
To benchmark algorithmic performance, a standardized series of tests were designed. Thirteen different programs of varying execution time and structure were implemented.
Assembly and executables were created from the source files. An arrival function was created which simulated task arrivals and deadlines over a specific point in time. A series of task request orders with a fixed arrival time were created and run against each algorithm to gauge algorithmic performance. Testing was completed on a single Intel $4$-core hyperthreaded i$7$ CPU. 

At the end of each request order run with a particular scheduling, a series of measurements were recorded. First, core measurements of temperature, speed, and utilization were taken at one second intervals throughout the entirety of the scheduling. Secondly, from the perspective of the tasks, individual task waiting time until being scheduled was recorded for each arrival in the system. Finally, all accumulation window sizes for the NPDW algorithm were stored.

\subsection{Code Samples} \label{code-samples}
Code samples and their assembly analysis for the tasks are described below. 
All code samples are implemented in C++20 and compiled with the GNU GCC compiler and default optimization levels on an Ubuntu 20.04 machine. The assembly for each program 
is produced with the \texttt{-S} flag provided by GCC. Many of the code samples are modified versions of the stress-ng~\cite{stress-ng} CPU benchmarking framework. 

\smallskip

\noindent
\textbf{1. Sieve of Eratosthenes:}
The Sieve of Eratosthenes algorithm is a classical algorithm used to determine the set of primes up until a given number $n$. A boolean array of size $n$ is created and each value initialized as true (except $1$ which is neither prime nor composite). If a number is deemed a prime, then all multiples from its square up until $n$ are deemed composite and the corresponding array value is set to false. This process repeats from $2$ to $\sqrt{n}$. 
The assembly produced by this code sample is quite diverse and covers a large portion of the assembly we analyze. Implementation made use of heavy comparison and jump instructions were also prominent. A single integer multiplication was utilized to perform the square operation,
and a series of bit shifts was utilized for the square root operation. 
Various addition and subtraction operations are performed throughout
the program.

\smallskip

\noindent
\textbf{2. Gray Codes:}  \label{gray-code}
A gray code is an alternate representation of the binary number system. Instead of the traditional representation in which the number is expressed through powers of 2, the gray code expresses the number such that consecutive numbers differ by only a single bit. 
The algorithm computes the gray encoding of a given number and stores it in its base-10 representation before converting the number back to its traditional base-10 representation. 
The produced assembly 
consists heavily of bitwise operations such as bitwise XOR, bitwise AND, and right bitshifts. Jump and comparison structures were prominent in the implementation of the looping structures of the code. 

\smallskip

\noindent
\textbf{3. Approximation of Pi:} \label{approx-pi}
The numeric constant PI can be obtained by an infinite series known as the Ramanujan-Sato Series:
$ {1}/{\pi} =$\\
$ ({2\sqrt{2}}/{9801}) \sum_{k=0}^\infty ({(4k)!}/{(k!^4)}) ({(26390k+1103)}/{(396^{4k})}).$\\
The produced assembly consists largely of floating point operations via the FPU register on $64$-bit numbers in addition to conditional jumps and movements necessary for the looping structure. 
Of all the code samples, the approximation of PI utilizes floating point operations the most heavily.

\smallskip

\noindent
\textbf{4. Recursive Fibonacci Algorithm:} \label{fib}
The Fibonacci series is 
defined as 
$f(0) = 0, f(1) = 1,$ and $f(n) = f(n-1) + f(n-2) \text{ for } n\ge 2$. 
The Fibonacci code tests the effects of heavy recursion on the scheduler. 

\smallskip

\noindent
\textbf{5. Naive Matrix Multiplication:} \label{matrix-multiply}
The matrix multiplication program computes the product of two matrices via the standard algorithm. 
Implementation of the algorithm allocates three matrices 
on the heap to assess the effects of heavy heap allocation and access on the scheduler. 
Double precision floating point numbers 
are stored in the matrices.
The resulting assembly consists of heap allocation call instructions, SIMD floating point operations, and a looping structure implemented with conditional jumps.
Compiler optimizations resulted in 
bitshift and bitwise operations in the resultant assembly. By the metrics obtained in Section~\ref{static}, the matrix multiplication test code is the most costly per instruction. 

\smallskip

\noindent
\textbf{6--13. Select Arithmetic Operations:} \label{arithmetic-ops}
The last sets of test codes perform a single arithmetic operation many times on a set of randomly generated operands. The arithmetic operations tested are addition, subtraction, multiplication, and division.
Eight executables are produced, each consisting of one of the four arithmetic operations and either double precision floating-point or integer values for the operands. Assembly for the executables are nearly identical with the exception being the opcode utilized for the different assembly operations. SIMD floating point instructions are utilized for double precision values, and the standard x86 arithmetic instructions are used for integer values. The goal of this test set is to isolate a single instruction and test relative performance differences in the scheduler when the single arithmetic instruction is changed.

\subsection{Results} \label{results}

In this section, we discuss the results obtained from experimentation of the four schedulers 
described in
Section~\ref{schedulings} with the thirteen task programs listed in 
Section~\ref{code-samples}.  

\subsubsection{Data Collection and Measurement} \label{collection}
Task arrivals were simulated and stored in input files, each of which contained a series of 2000 scheduled tasks with an arrival time and ideal deadline. Arrival times were randomly generated via the Poisson distribution and one of the thirteen programs was selected at random for each arrival. With this randomization scheme, five task arrival input files were generated which were run on each scheduler. 
During execution, collection scripts monitored and recorded CPU core utilization, speed, and temperature data. Additionally, waiting times per task and window resizings (in the case of NPDW) were recorded by the system.
The collected measurements were utilized to calculate the 
load distribution, temperature distribution, and scheduling fairness. 
The measurements and corresponding metrics 
are recorded for each scheduling benchmark test for the schedulers described in
Section~\ref{schedulings}. Between each individual scheduler test, the CPU was given ten minutes of idle time (no task execution apart from necessary system processes) to fully recuperate.

\subsubsection{Accumulation Window Size Resizing} \label{accumulation-stabilization}
\begin{figure*}[!t]
\centering
\subfloat[Trial 1]{\includegraphics[width=1.8in]{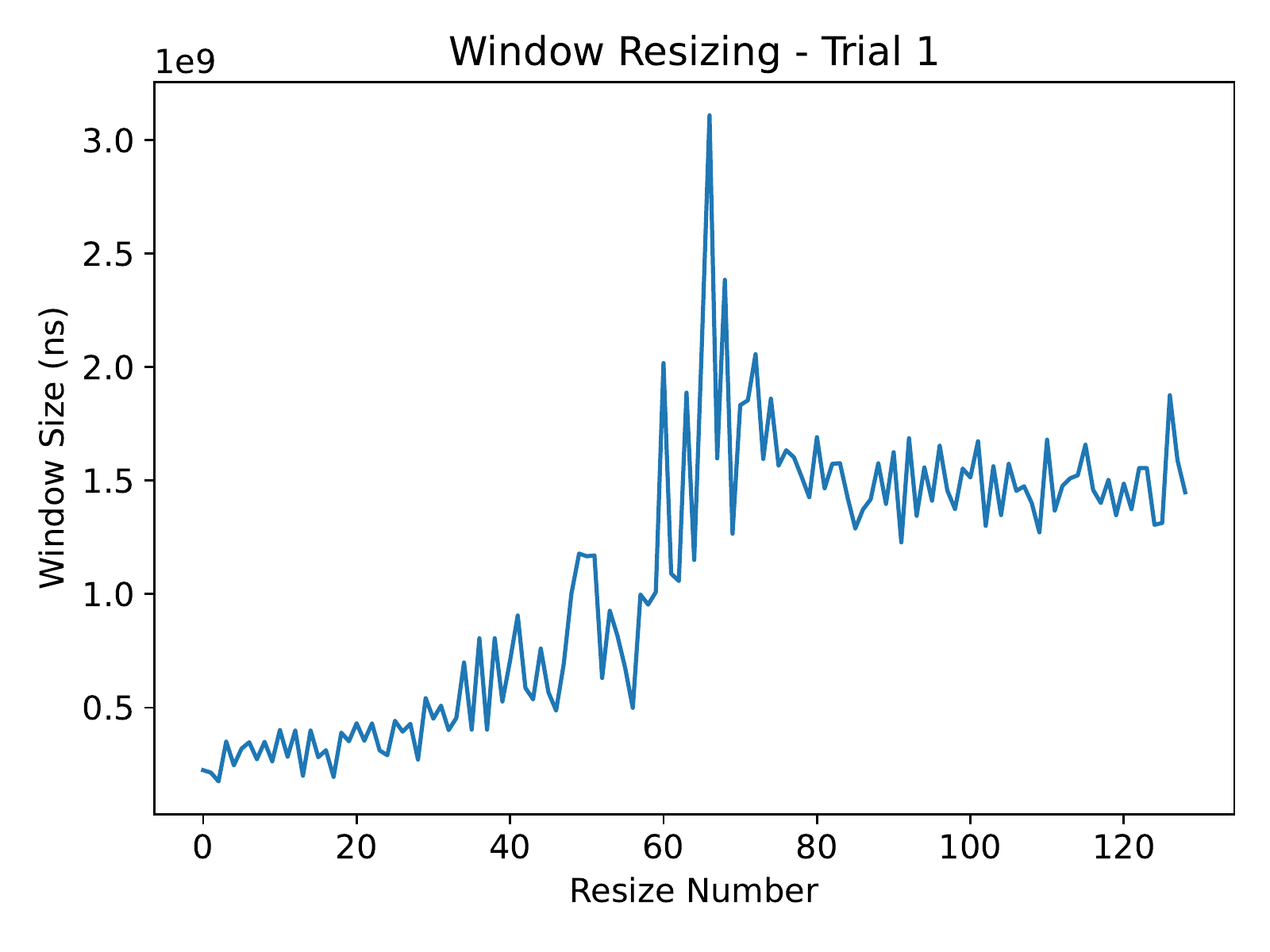}%
\label{window_resize_trial1}}
\subfloat[Trial 2]{\includegraphics[width=1.8in]{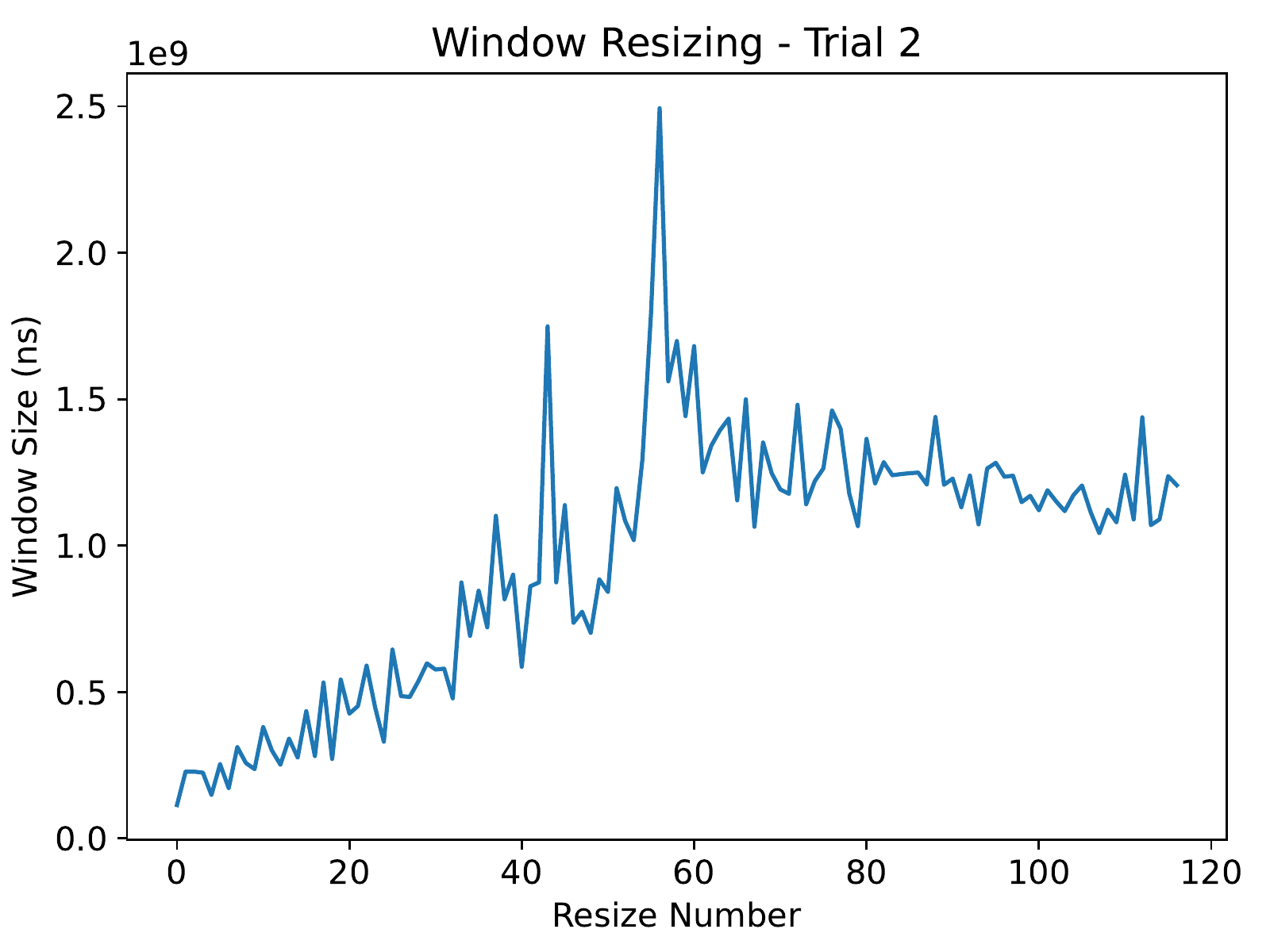}%
\label{window_resize_trial2}}
\subfloat[Trial 3]{\includegraphics[width=1.8in]{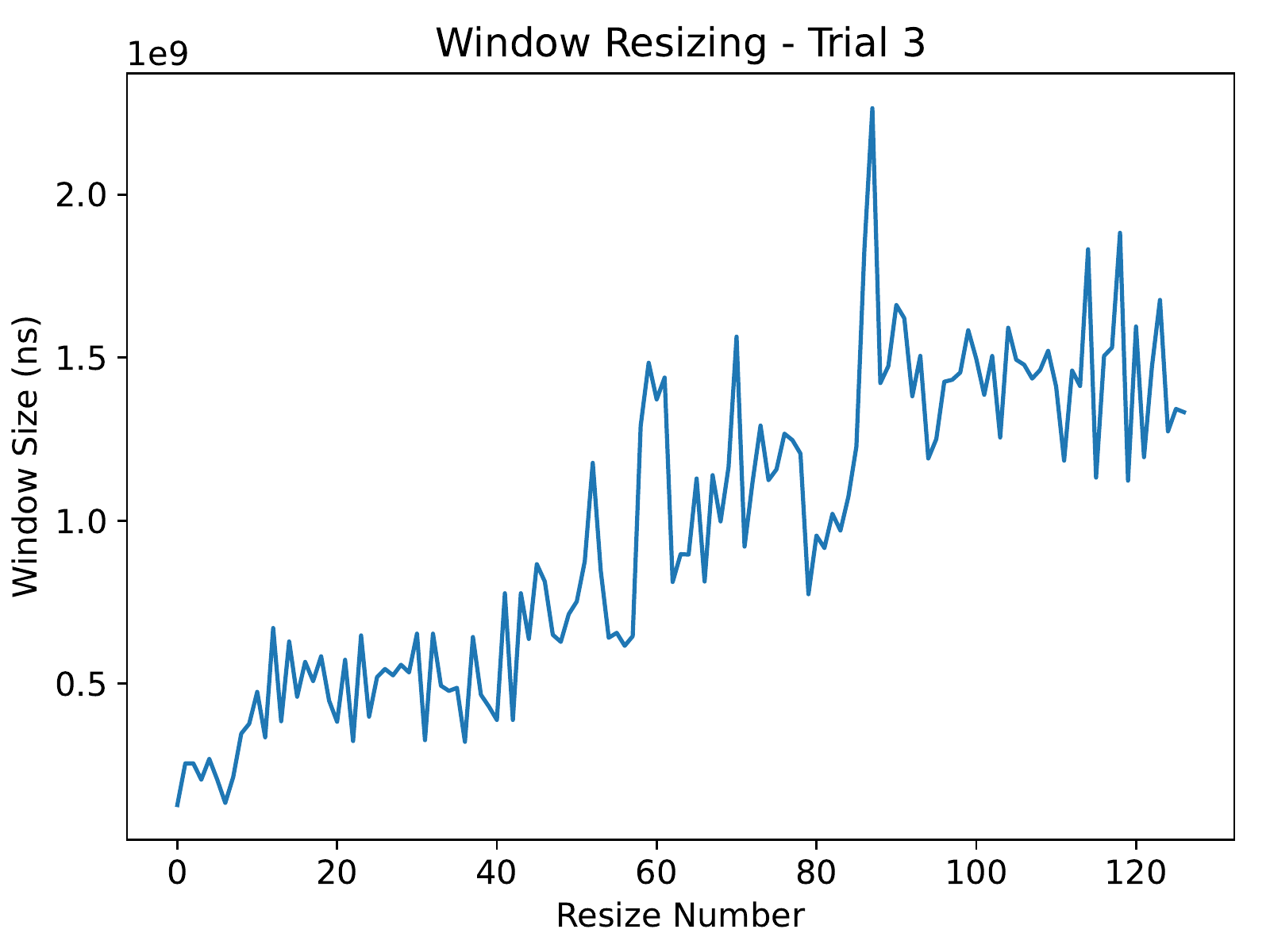}%
\label{window_resize_trial3}}
\subfloat[Trial 5]{\includegraphics[width=1.8in]{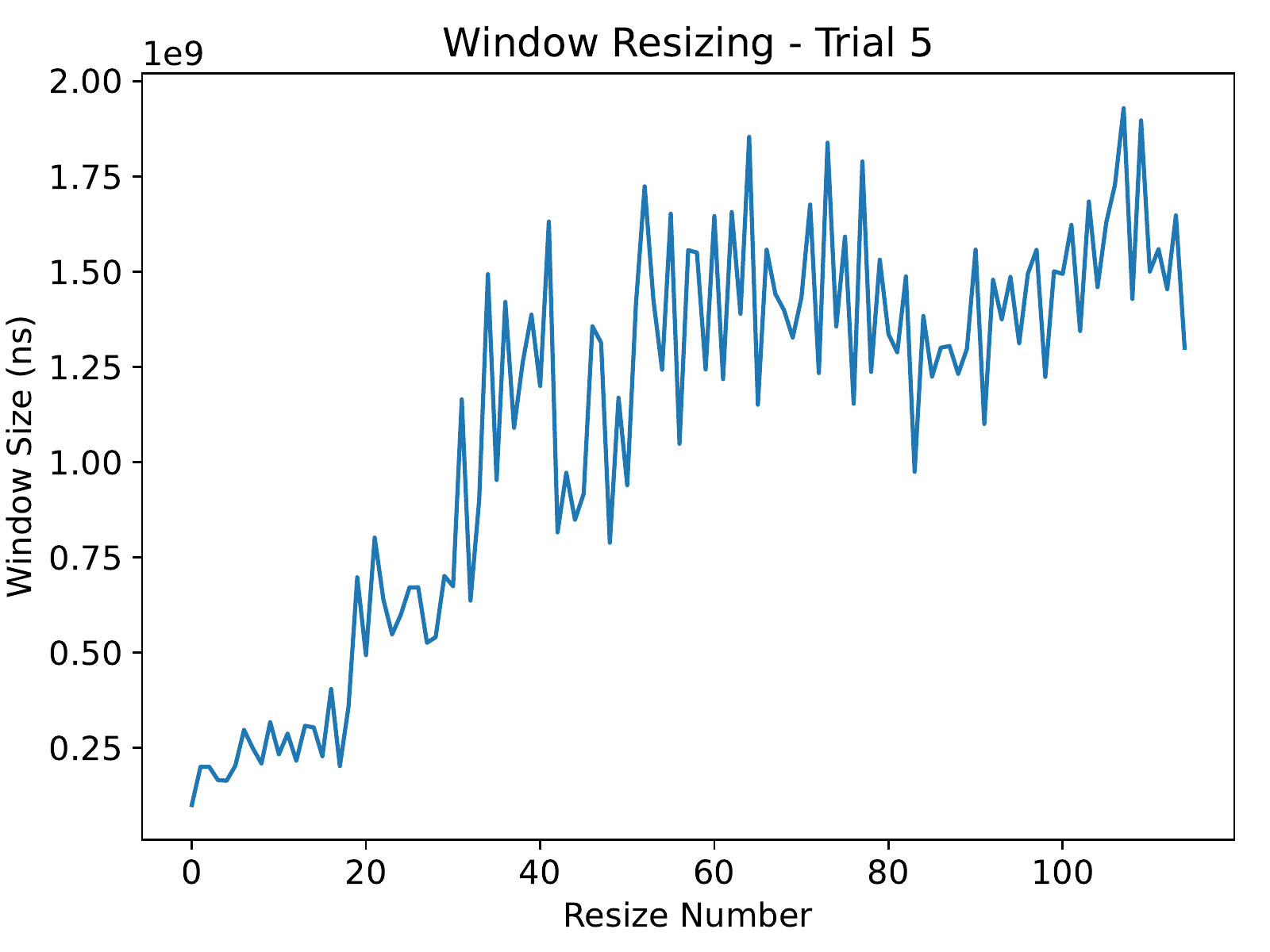}
\label{window_resize_trial4}}
\caption{Accumulation window resizing trials. (Only four trials results are shown due to space limit.)}
\label{fig:accumulation-graph}
\end{figure*}

Within our scheduling algorithm, we assert that under randomized task arrival conditions, the accumulation window size will reach a state of stabilization in which resizings center around a timeframe. This claim is empirically tested by measuring accumulation window size throughout the five random task arrivals generated via a Poisson Process. The results are shown in Fig.~\ref{fig:accumulation-graph}.

All five trials show that the window resizing happens within the range of [1, 2] seconds. This trend is particularly apparent in trials one and two. Stabilization within this time window range indicates that the scheduler has approximated an ideal 
accumulation window in which the tradeoffs among task waiting during
accumulation, 
temperature and load distribution of cores, and 
core utilization can be balanced.

\subsubsection{Load Distribution and Core Utilization} \label{load-distribution}
An important feature in our scheduler's design is the element of cooperation among the cores. 
As discussed in Section~\ref{matching}, our scheduler promotes equitable distribution of load among the 
cores through the bipartite matching.
Distribution of load is measured using the standard deviation of average core utilization among all cores throughout the scheduling. An ideal scheduling would fully equalize load distribution among 
cores and utilization is effectively balanced throughout a scheduling test set. Therefore, the closer the obtained metric is to $0$, the more optimal the load balancing.
The results of the scheduler load balancing assessments are shown in Table~\ref{tbl:load}.
It is observed 
that our NPDW scheduler exhibits greater load distribution via standard deviation of average load compared to other schedulers, indicating  
that the CPU load is better managed via our scheduler compared to baseline scheduling algorithms.

\vspace*{-10pt}
\begin{table}[H]
\caption{Standard Deviation of Average Core Load ($\%$)}
\label{tbl:load}
\centering
\vspace*{-5pt}
\begin{tabular}{|c|c|c|c|c|}
\hline
Trial Number & NPDW & FCFS & SJF & EDF \\
\hline
1 & 17.5647 & 24.9386 & 25.3857 & 25.3857 \\
\hline
2 & 18.0694 & 24.4169 & 23.4498 & 24.0413 \\
\hline
3 & 17.7721 & 23.3477 & 24.2235 & 23.9290 \\
\hline
4 & 18.8619 & 25.0606 & 24.2586 & 24.9282 \\
\hline
5 & 19.0787 & 21.3378 & 21.9833 & 20.9025 \\
\hline
\end{tabular}
\end{table}

Alongside the load distribution metric, the raw utilization measurements defined as utilization throughput are 
measured and shown in Table~\ref{tbl:usage}. 
In NPDW, in order to obtain optimal scheduling batches through the bipartite matching, time windows are established. Consequently, a certain amount of utilization 
is potentially sacrificed in order to perform a well-informed scheduling at a later time,
allowing the processor cores to cool down. This sacrifice
of utilization is noticed in Table~\ref{tbl:usage} when 
compared to the baseline scheduling algorithms which lack this feature. 
These measurements
showcase the tradeoff between the level of utilization our scheduler sacrifices and the better managed core temperature and load distribution promoted by our scheduler.
\vspace*{-10pt}
\begin{table}[H]
\caption{Average Utilization Throughput ($\%$)}
\label{tbl:usage}
\centering
\vspace*{-5pt}
\begin{tabular}{|c|c|c|c|c|}
\hline
Trial Number & NPDW & FCFS & SJF & EDF \\
\hline
1 & 56.6090 & 69.1085 & 70.9907 & 68.9736 \\
\hline
2 & 57.2318 & 61.8277 & 61.8455 & 61.0444 \\
\hline
3 & 60.1866 & 65.6169 & 67.1898 & 66.8205 \\
\hline
4 & 58.3795 & 69.2843 & 69.5688 & 70.0171 \\
\hline
5 & 57.1683 & 58.2924 & 59.0817 & 58.4811 \\
\hline
\end{tabular}
\end{table}

\subsubsection{Temperature Distribution and Control} \label{temperature-distribution}
In correlation with load distribution, core temperature distribution and control is a desired goal of the scheduler. To ensure full core operability,
our NPDW scheduler reduces load 
on high-temperature cores and increases the load 
on low-temperature cores. The desired aim is to unilaterally raise the temperature of all cores together as they are stressed through a benchmark test set. Throughout the scheduling, the standard deviation for each core's temperature 
is measured. 

Raw temperature values 
are recorded across the scheduling in order to determine the ability of the scheduler to control the CPU's temperature to maintain optimal core functioning and to prevent mandated thermal throttling in the hardware.
To assess the distribution of temperature across the different CPU cores in scheduling, the standard deviation of temperature measurements across a test schedule set were taken and averaged. 
\vspace*{-10pt}
\begin{table}[H]
\caption{Thermal Distribution ($^\circ$C)}%
\label{tbl:thermal}
\centering
\vspace*{-5pt}
\begin{tabular}{|c|c|c|c|c|}
\hline
Trial Number & NPDW & FCFS & SJF & EDF \\
\hline
1 & 3.1101 & 3.3204 & 3.3928 & 3.2114 \\
\hline
2 & 3.1473 & 3.5698 & 3.5251 & 3.5572 \\
\hline
3 & 3.2611 & 3.3134 & 3.3663 & 3.4279 \\
\hline
4 & 3.1480 & 3.4035 & 3.3023  & 3.4466 \\
\hline
5 & 3.0745 & 3.7743 & 3.6908 & 3.8383 \\
\hline
\end{tabular}
\end{table}
In assessing the thermal distribution of the various schedulers, it is observed that our NPDW algorithm
outperforms 
all other scheduling algorithms in all five trials conducted implying greater overall power and energy conservation by the processor. 

In addition to assessing the thermal distribution of the processor cores, the overall processor temperatures were measured.
Table~\ref{tbl:temperature}
showcases the CPU temperature measurements.
\vspace*{-10pt}
\begin{table}[H]
\caption{Average Temperature Throughout Scheduling ($^\circ$C)}
\label{tbl:temperature}
\centering
\vspace*{-5pt}
\begin{tabular}{|c|c|c|c|c|}
\hline
Trial Number & NPDW & FCFS & SJF & EDF \\
\hline
1 & 61.7287 & 64.5572 & 64.6730 & 63.9441 \\
\hline
2 & 60.8131 & 63.1920 & 63.2756 & 62.9305 \\
\hline
3 & 61.5669 & 63.3173 & 63.8388 & 63.3883 \\
\hline
4 & 60.3174 & 64.1677 & 63.8233 & 63.9758 \\
\hline
5 & 59.3445 & 61.2127 & 61.4994 & 61.0850 \\
\hline
\end{tabular}
\end{table}
Overall observed processor temperature throughout the scheduling is lower in NPDW
than in other scheduling algorithms. The results demonstrate that the introduction of cooldown periods via time windows in our scheduler effectively lower CPU core temperature throughout. 

\begin{figure*}[!h]  
\centering
\subfloat[Trial 1]{\includegraphics[width=1.8in]{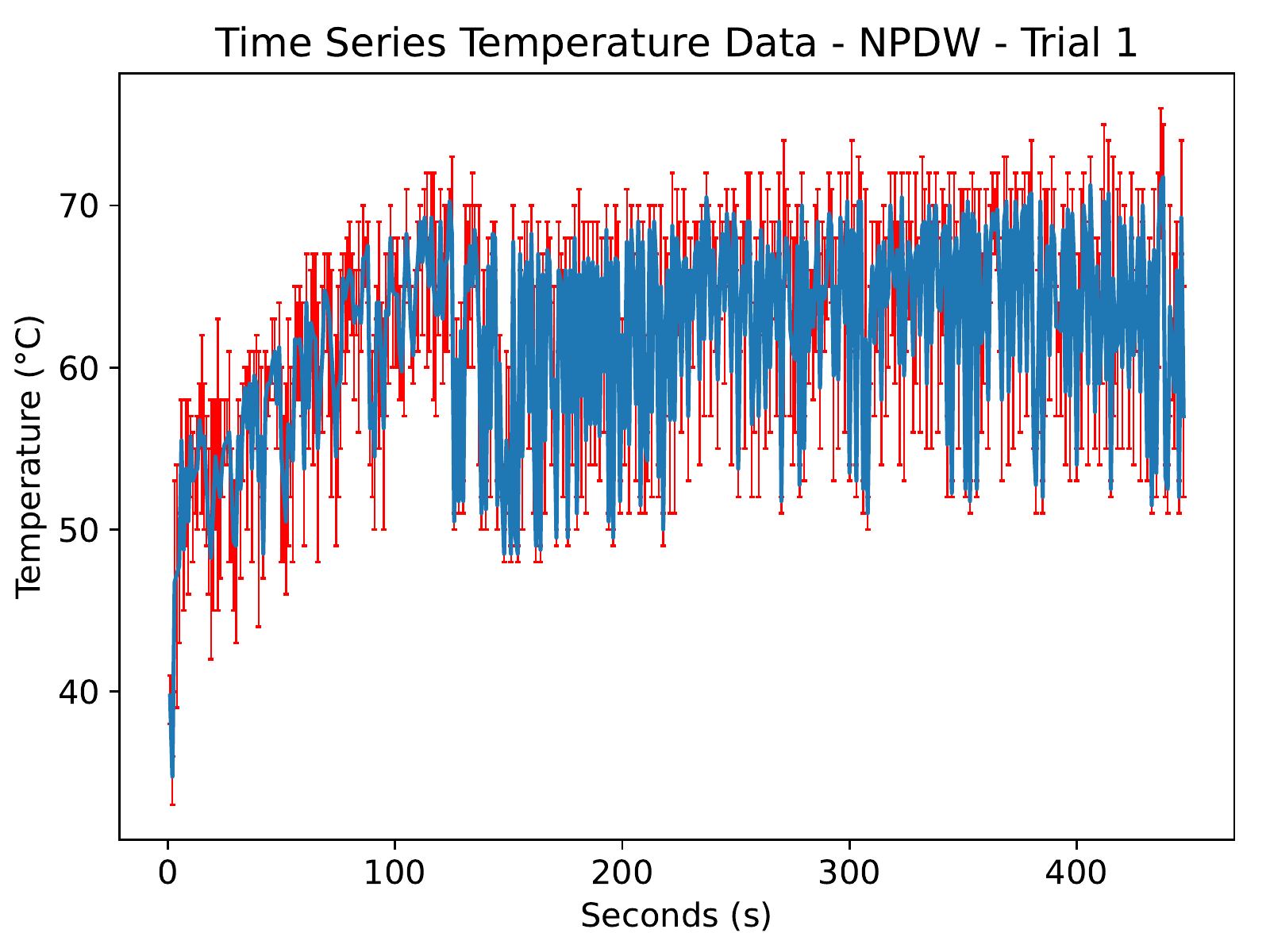}%
\label{temp_err_trial1}}
\subfloat[Trial 2]{\includegraphics[width=1.8in]{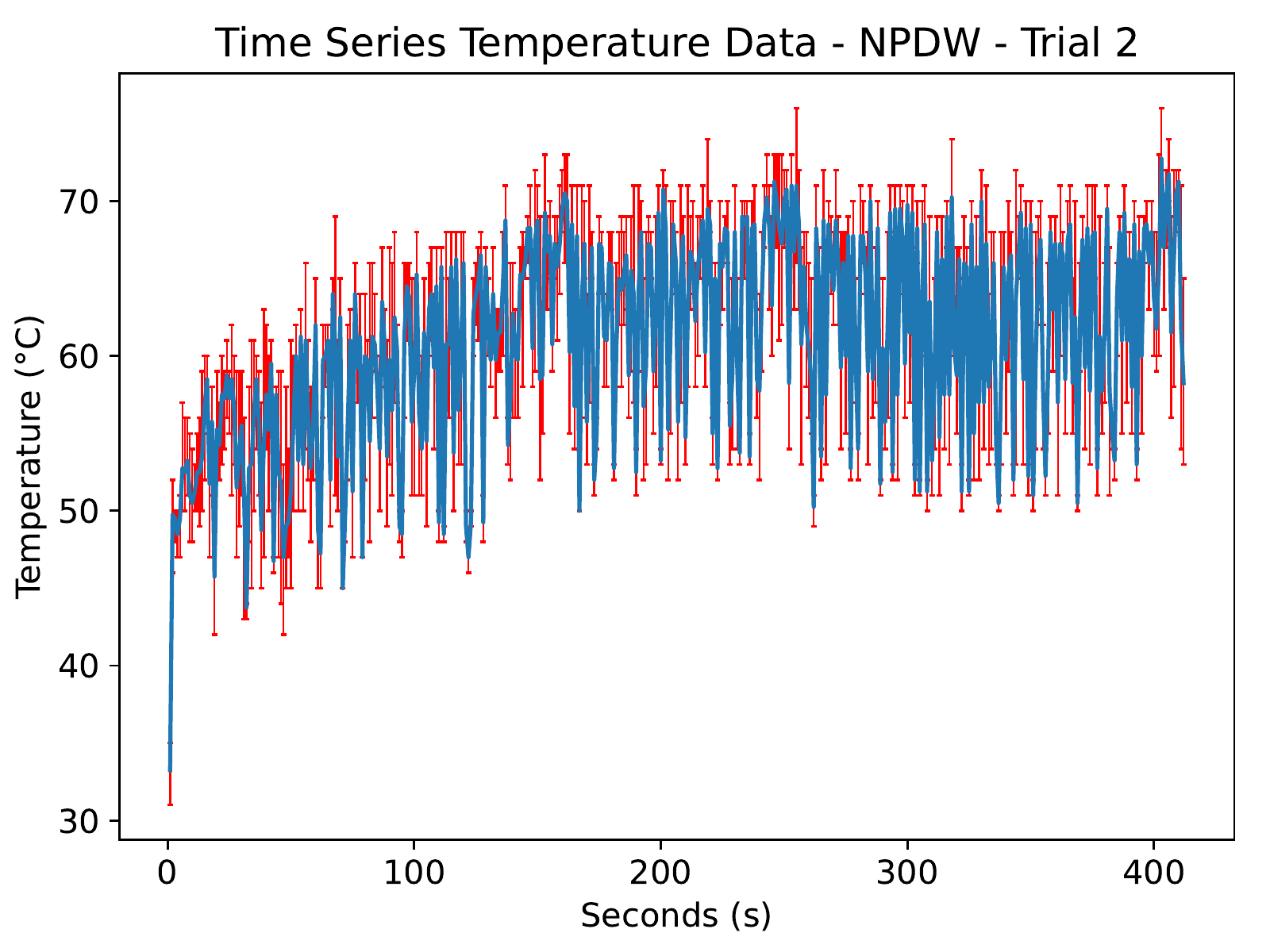}%
\label{temp_err_trial2}}
\subfloat[Trial 3]{\includegraphics[width=1.8in]{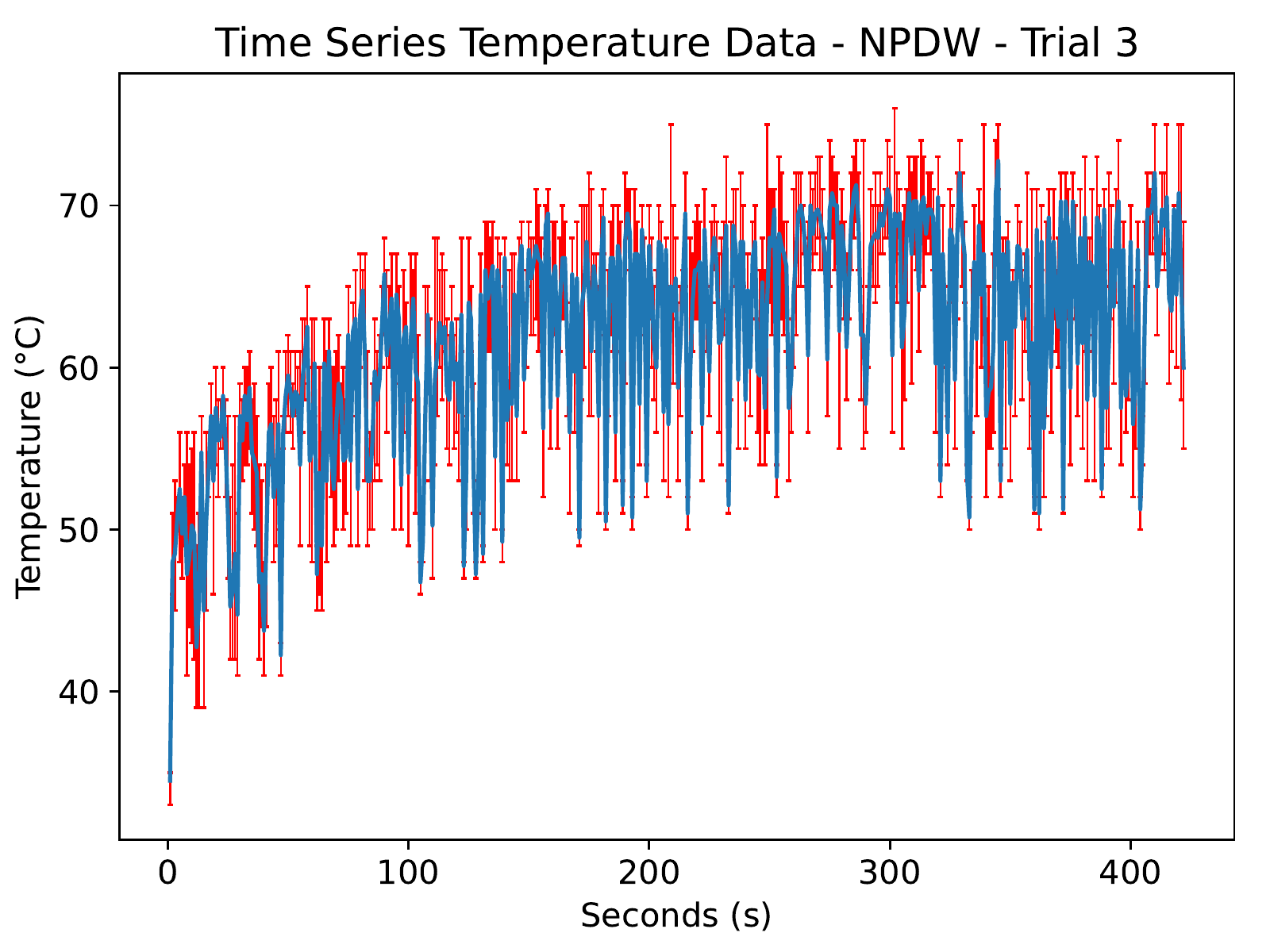}%
\label{temp_err_trial3}}
\subfloat[Trial 5]{\includegraphics[width=1.8in]{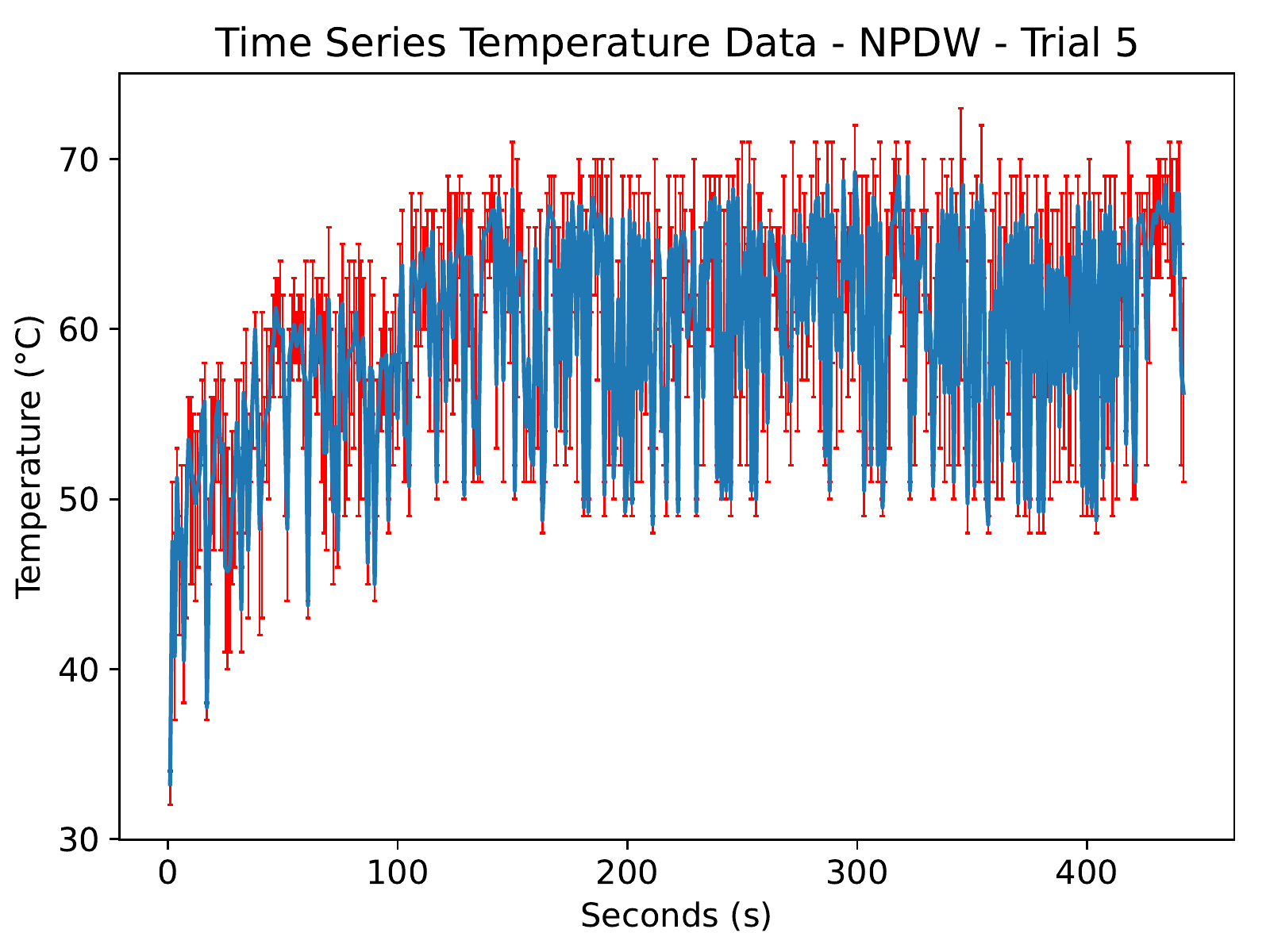}
\label{temp_err_trial4}}
\caption{Temperatures. (Only four trials results are shown due to space limit.)}
\label{fig:temp-error-bars}
\end{figure*}

Fig.~\ref{fig:temp-error-bars} demonstrates CPU temperature throughout scheduling after a state of equilibrium is achieved. The error bars indicate min and max core temperatures at a given timeframe.
CPU temperature balancing shown in Fig.~\ref{fig:temp-error-bars} coincides with the stabilization of time window size shown in Fig.~\ref{fig:accumulation-graph}. 

\subsubsection{Scheduling Fairness} \label{scheduling-fairness}
Relative fairness from the perspective of individual tasks is an important metric in scheduling.
To assess our scheduler's credit based system of fairness, a metric of fairness is recorded. The overall scheduler fairness 
is defined as the coefficient of variance of the waiting times of tasks throughout the system. The coefficient of variance is utilized as opposed to standard deviation in order to appropriately account for differences in waiting time means that exist between schedulers. By normalizing with the mean, the relative effect of deviations from the mean are appropriately considered. This measurement 
is taken across the entire set of scheduled tasks after execution to gauge fairness measurements between the schedulers. The 
results are shown in Table~\ref{tbl:wait-time}.
\vspace*{-10pt}
\begin{table}[H]
\caption{Coefficient of Variance of Task Waiting Time}
\label{tbl:wait-time}
\centering
\vspace*{-5pt}
\begin{tabular}{|c|c|c|c|c|}
\hline
Trial Number & NPDW & FCFS & SJF & EDF \\
\hline
1 & 0.8767 & 2.2133 & 3.5449 & 3.9664 \\
\hline
2 & 2.8039 & 2.6982 & 3.9802 & 4.2299 \\
\hline
3 & 0.9863 & 3.0001 & 3.9915 & 5.2813 \\
\hline
4 & 1.0283 & 2.0739 & 4.8250 & 4.7406 \\
\hline
5 & 1.1709 & 4.0869 & 4.6766 & 5.2344 \\
\hline
\end{tabular}
\end{table}
NPDW outperforms the baseline schedulers based on the 
measurement of scheduling fairness. Based upon average coefficient of variance of task waiting time, the relative deviation between task schedules in NPDW is lower than that of alternative schedulings. As a result of the inclusion of task credit and accumulation time windows into the scheduling, the scheduler is appropriately able to introduce an element of fairness that is nonexistent with the other schedulers. Second to NPDW in performance was FCFS followed by SJF and EDF. 

\section{Conclusions} \label{conclusions}
In this paper, we proposed a novel approach to multicore system scheduling
known as Non-Preemptive Dynamic Windowing (NPDW) for effective load and temperature
balancing. We introduced the concept of dynamic time windows 
where an accumulation window is defined as a period of waiting time in which 
a group of tasks are collected followed by an execution window in which tasks are 
executed on CPU cores. We described how core centric measurements such as load, 
speed and temperature were utilized alongside task centric measurements such as 
credit, execution categorization, and deadline to create
the concept of matching affinity. Furthermore, we introduced how a bipartite graph 
representation can be utilized with matching affinity to apply a modified
Gale-Shapely algorithm to create task-core matchings to be applied during 
execution windows. Lastly, the metrics of task and window performance were 
defined to create a window resizing heuristic to drive the creation of the 
next time window size based on the current and previous window sizes.

To substantiate formulations given in the paper, we provided the proofs of
credit spending fairness, task to core matching affinity, optimality of core to task matching affinity,
window resizing direction, and the next window size
heuristic. Furthermore, we conducted experimentation to assess the performance 
of our scheduler for task arrival scenarios randomized with the Poisson process. 
The experimentation results confirm that our NPDW scheduler outperforms 
common baseline schedulers
in terms of load distribution, thermal distribution, and fairness at the expense 
of overall utilization and total task runtime. Moreover, we 
observed that relative 
stability of the time window size was achieved in long-term running of NPDW 
with randomized arrivals. 

In future work, we wish to expand experimentation of our scheduler
onto larger multicore environments. Due to hardware limitations, we could only test
on a quad-core Intel CPU with eight hardware threads. Since our scheduler 
strongly considers individual core load and temperature when performing task-to-core 
matching, we believe significant thermal and load balancing benefit larger than 
what was observed in this paper can be achieved on high performance multicore 
systems. 

In modern computing, thermal control and power efficiency are of 
major significance from both ecological and economic perspectives. The need to 
develop systems which exploit multi and many core architectures to 
reduce wasteful heat generation is of utmost importance in compute intensive 
applications. We believe that our approach utilized in the NPDW scheduler, 
emphasizing processor core consideration of temperature and window based
scheduling may be utilized in a wide array of systems to effectively distribute
heat and reduce overall system temperature.

\bibliographystyle{IEEEtran}
\bibliography{mmichel-hlee-NPDW}


\begin{IEEEbiography}[{\includegraphics[width=1in,height=1.25in,clip,keepaspectratio]{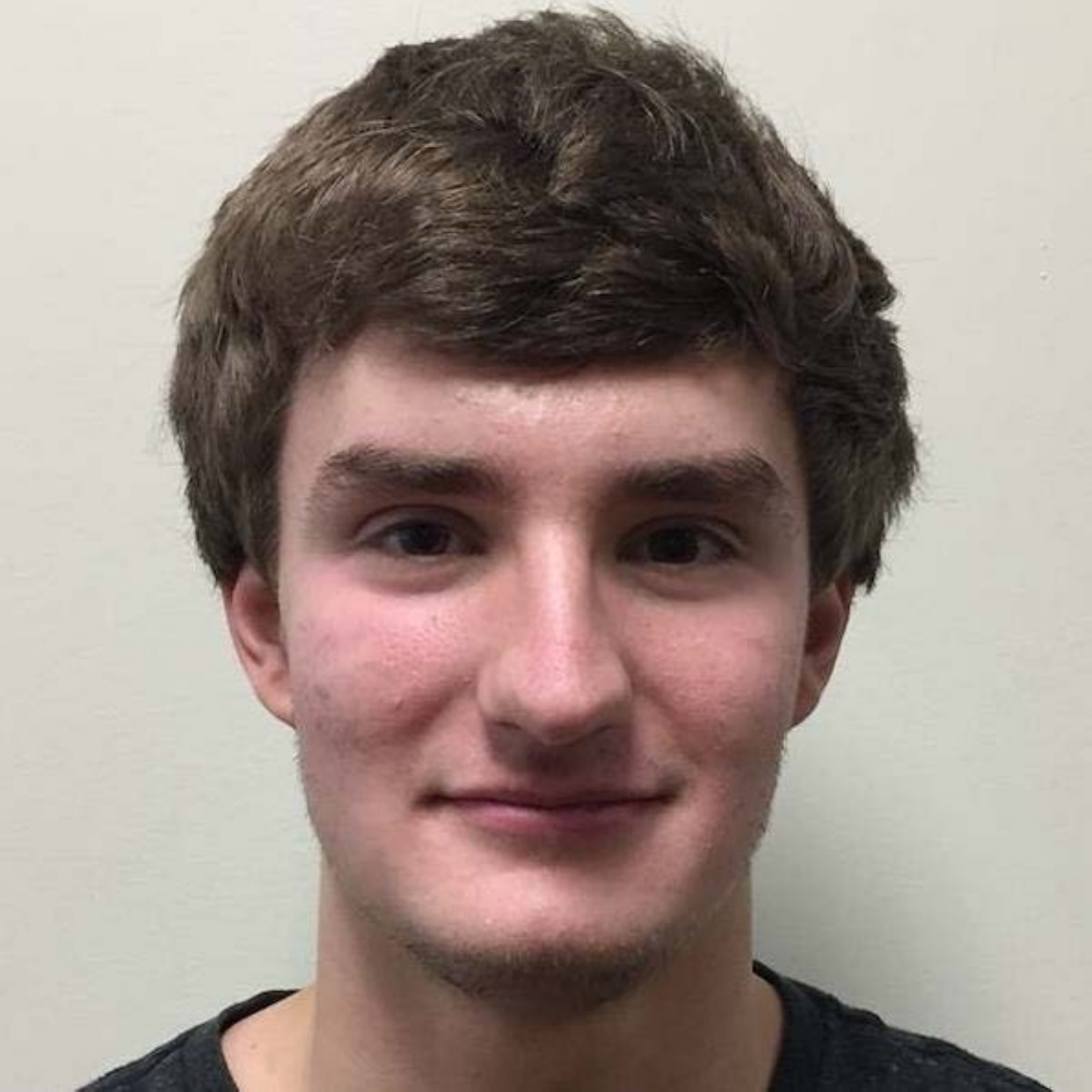}}]{Matthew Michel} is an undergraduate student in the Department of Computer Science 
and Engineering at Texas A\&M University in College Station, Texas. 
He has gained experience as an intern working on modern C++ scientific applications
within the areas of mixed-integer programming, constraint optimization, and graph-based models.
Additionally, he has worked on cloud-based computing and automation systems, 
big-data analytics, and GraphQL API development. 
His areas of interest include multicore processor scheduling, high
performance computing, and functional programming.
\end{IEEEbiography}

\begin{IEEEbiography}[{\includegraphics[width=1in,height=1.25in,clip,keepaspectratio]{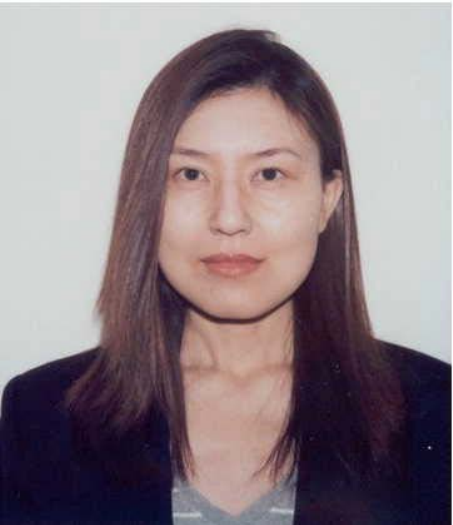}}]{Hyunyoung Lee} is a faculty member in the Department of Computer Science 
and Engineering at Texas A\&M University in College Station, Texas. 
She received her Ph.D.\ from Texas A\&M University in 2001, her M.A.\ from Boston University 
in 1998, and her M.S.\ and B.S.\ both from Ewha University in South Korea
in 1992 and 1987, respectively.
Dr.\ Lee has also worked as a faculty member in the Computer Science department at the 
University of Denver.
Her research focus is on distributed computing, in particular, energy-conscious programming 
models and scheduling in multicore systems, dynamic distributed systems, and
distributed shared memory consistency models.
\end{IEEEbiography}
\vfill

\end{document}